\documentclass[usenatbib]{mn2e}

\usepackage{amssymb}
\usepackage{times}
\usepackage{epsfig}
\usepackage{colordvi}

\newcommand{\be}{\begin{equation}}
\newcommand{\ee}{\end{equation}}
\newcommand{\ls}{\ell_\mathrm{s}}
\newcommand{\lh}{\ell_\mathrm{h}}

\newcommand{\lnth}{\ell_\mathrm{nth}}
\newcommand{\lhls}{\ell_\mathrm{h}/\ell_\mathrm{s}}
\newcommand{\msun}{{{\rm M}_{\sun}}}

\newcommand{\rg}{R_\mathrm{g}}

\newcommand{\tin}{kT_\mathrm{max}}
\newcommand{\ktbb}{\tin}
\newcommand{\taut}{\tau}
\newcommand{\taub}{\tau_\mathrm{p}}
\newcommand{\kte}{kT_\mathrm{e}}
\newcommand{\ktseed}{kT_\mathrm{seed}}
\newcommand{\nh}{N_\mathrm{H}}
\newcommand{\betac}{\beta_\mathrm{c}}
\newcommand{\ginj}{\Gamma_\mathrm{inj}}

\newcommand{\Refl}{R}
\newcommand{\sigmat}{\sigma_\mathrm{T}}
\def\Ginga{{\it Ginga}}
\newcommand{\xte}{{\it RXTE}}
\def\gro{{\it CGRO}}
\def\asca{{\it ASCA}}
\def\sax{{\it Beppo\-SAX}}
\def\me{m_{\rm e}}
\def\xg{X/$\gamma$}

\title[Broad-band spectra of Cyg X-1]{Broad-band spectra of Cyg X-1 and
correlations between spectral characteristics}

\author[A. Ibragimov et al.]
{Askar Ibragimov,$^{1,2}$\thanks{E-mail:
askar.ibragimov@oulu.fi (AI), juri.poutanen@oulu.fi (JP)}
Juri Poutanen,$^1$\footnotemark[1]\thanks{Corresponding Fellow, NORDITA, Copenhagen}
Marat Gilfanov,$^{3,4}$ Andrzej A. Zdziarski,$^5$ \newauthor and Chris R. Shrader$^6$ \\
$^1$Astronomy Division, PO Box 3000, FIN-90014 University of Oulu, Finland\\
$^2$Kazan State University, Astronomy Department, Kremlyovskaya 18, 420008
Kazan, Russia \\
$^3$Max-Planck-Institut f\"ur Astrophysik, Karl-Schwarzschild-Str. 1, 85740
Garching, Germany\\
$^4$Space Research Institute, Russian Academy of Sciences, Profsoyuznaya 84/32,
117810 Moscow, Russia\\
$^5$Centrum Astronomiczne im.\ M. Kopernika, Bartycka 18, 00-716 Warszawa, Poland \\
$^6$Laboratory for High-Energy Astrophysics, NASA Goddard Space Flight Center,
MD 20771 Greenbelt, USA
}

\date{accepted, received}

\pagerange{\pageref{firstpage}--\pageref{lastpage}}
\begin{document}


\maketitle
\label{firstpage}

\begin{abstract}
We present the results of spectral analysis of 42 simultaneous
broad-band \Ginga--OSSE and \xte--OSSE observations of Cyg X-1
carried out in 1991 and 1996--1999. The broad-band  spectra  from 3 to
$\sim 1000$ keV   can be well described by thermal Comptonization model
with reflection from the cold disc, with an additional soft component
visible below 10 keV.
  The relative contribution of this
component to the total energy flux appears to be higher in the
spectra with larger reflection amplitude and steeper photon index
of the thermal Comptonized component. We consider a number of
physically realistic models to describe the shape of the $E\la 10$
keV excess. The additional soft component can result from thermal
Comptonization by electrons with a low Compton parameter, or can
be a part of a nonthermal, power-law like emission extending above
1 MeV.

We study correlations between parameters obtained from the spectral
fits with different models. We confirm a general correlation
between the photon index $\Gamma$ and the amplitude of reflection
$\Refl$.
We find that simple phenomenological models (like power-law plus
Compton reflection) applied to the narrow band (3--20 keV) data
overestimated the values of $R$ and $\Gamma$, although
the simple models did rank correctly the
spectra according to $\Refl$ and $\Gamma$, as it was demonstrated in the
original  publications on this subject.

The dynamic corona model provides a satisfactory description of
the observed correlation, while the hot inner disc models have
problems in reproducing it quantitatively. On the other hand,
in the context of the dynamic corona model it is difficult to
understand correlations with the timing characteristics, which
seems natural in the hot disc scenario. We do not find significant correlation between the
electron temperature and other spectral parameters, while the
optical depth of the hot medium seems to decrease when the
spectrum becomes softer. It is also shown that spectral
parameters are well correlated with the timing characteristics of
the source.
\end{abstract}


\section{INTRODUCTION}

Matter accreting onto a black hole, whether supermassive in Seyfert galaxies or
stellar mass in Galactic X-ray binaries, releases most of its gravitational
energy in the form of X-rays deep in the potential well. Accretion may proceed
in a number of regimes. The accreting gas can approach the black hole in a
disc-like configuration \citep{ss73} if the gravitational energy is effectively
transported away in the form of radiation, or in a form of an almost spherical
flow if the energy exchange mechanism between protons (carrying most of the
energy) and electrons is inefficient \citep*{sle76,ich77,nmq98}. Magnetic
fields can play an important role transporting large fraction of the total
available energy and dissipating it in a rarefied medium (corona) above the disc
\citep*{grv79,tp92,sz94,bel99,ms00}. However, every model is based on a number
of assumptions that include prescriptions for the viscosity, the vertical
distribution of the energy release through the flow, the energy transport
mechanisms, etc. Given the difficulties in the accretion physics, observations
should help in choosing among the different possibilities as well as guide
theoreticians in the right direction.

Cygnus X-1, one of the best studied black hole binaries (BHB), served as an
accretion disc laboratory since the end of the 1960's. The most dramatic
observed phenomena are the spectral state transitions  occurring every few
years, when the source, typically emitting most of its energy at about 100 keV in
the hard state, switches to a   soft state consisting of a prominent $\sim 1$
keV black-body and a power-law-like tail. The
hard state spectrum was believed to originate from thermal Comptonization in a
hot electron cloud \citep*{sle76,ich77,st79,st80}. The black-body looking
soft-state spectrum was associated with the optically thick accretion disc
\citep{ss73}, while the origin of $\sim100$ keV emission in that state was not
discussed much, because the detailed spectrum was not available.

During the last decade, the quality of the \xg-spectra increased dramatically
improving our knowledge as well as producing many new questions. In the hard
state, the spectrum turns out to be rather complicated containing a number of
components. Thanks to the broad-band coverage by \Ginga\/ and \gro\/ (and later
by \asca, \xte\/ and \sax), and advances in modelling of Comptonization at
mildly relativistic temperatures \citep{c92,ps96}, the parameters of the
electron cloud where Comptonization takes place were determined to a high
accuracy. In Cyg X-1, the electron temperature  of $\kte\simeq 100\pm50$ keV and
Thomson optical depth of $\taut\simeq 1-2$ were found to be typical
\citep*{zdz96,zdz97,gier97,p98,ds01,fro01,aaz04}. Compton reflection bump, a signature
of the presence of cold matter in the vicinity of the X-ray emitting source, was
discovered \citep{done92,ebis96,gier97}. The black body associated with the
cooler accretion disc \citep{bh91,bc95,ebis96} and an additional soft excess at
a few keV of unknown origin further complicate spectral decomposition
\citep{ds01,fro01}.  A high energy excess at $> 500$ keV discovered by \gro\/
\citep{mccon94,ling97} gives some clues about presence of nonthermal particles
in the source.

The \xg-ray soft state spectrum has been studied extensively by simultaneous
observations with \asca, \xte, \sax, and \gro\ during summer 1996.  In addition
to the dominating black-body, a long power-law like tail extending up to 10 MeV
was discovered \citep{mccon02}. The high-energy spectrum could be well described
by single Compton scattering off electrons having   a nearly power-law
distribution (\citealt{pc98,p98}; \citealt{gier99}, hereafter G99; \citealt{fro01}).
The Compton reflection was stronger
than in the hard state, which was interpreted as a change in the geometry of the
system, from the hot inner flow in the hard state to the standard Shakura-Sunyaev
disc with a nonthermal corona in the  soft
state (\citealt*{bkb77};  \citealt{ich77}; \citealt*{pkr97, lm97}; \citealt{esin98,pc98}).
Evaporation and condensation of the gas can
provide a physical basis for the change of the transition radius between
standard disc and hot inner flow \citep*{mlm00,rc00}. A smaller reflection
fraction in the hard state, however, can be explained by a beaming of the
primary emission away from the disc due to mildly relativistic motion of the
emitting plasma (ejection model, see \citealt*{bel99,bel99b,mbp01}, hereafter MBP01).
In order to distinguish among different possibilities, one needs to compare model
predictions with other observational facts.

A few, well separated in time, broad-band spectra do not give us
a good picture about relations between different components such
as, for example, Comptonized continuum and the reflection bump.
One can study these relations on a larger data set in a narrower
energy band. \citet*{zls99} (hereafter ZLS99) and
\citet*{gcr99} (hereafter GCR99) analyzing data from \Ginga\/ and
\xte, respectively, showed that the photon spectral slope of the
Comptonized emission, $\Gamma$, is strongly correlated with the
amplitude of Compton reflection component,
$\Refl=\Omega/2\pi$, where $\Omega$ is the solid angle
the cool material covers as viewed from the source of primary
X-rays. This correlation exists for individual BHBs and Seyfert
galaxies as well as in a sample of sources
\citep*[see also][]{gcr00,rgc01}. \citet{zdz03} studied possible
statistical and systematic effects and concluded that the
correlation exists beyond any reasonable doubts. Similar
correlation also exists for Fourier-frequency resolved spectra,
i.e., those corresponding to the variability in a given range of
Fourier frequencies \citep*{rgc99}.

The observed correlation provides extremely important clues to the geometry of
the accreting material and can be used for testing theoretical model.   The fact
that $\Gamma$ and $\Refl$   are correlated is a natural consequence of
co-existence of the cold media (accretion disc) and a hot Comptonizing gas in
the vicinity of the black hole. The cold material acts as a source of seed
photons for Comptonization and, at the same time, reflects and reprocesses the
hard radiation produced in the hot gas.

The cold disc with the hot inner flow model naturally produces
the correlation if there is an overlap between hot and cold
phases (\citealt{pkr97}; ZLS99). However, the $\Gamma$--$\Refl$
dependence observed in BHBs can be quantitatively reproduced only if the ratio
of the seed photon temperature, $\ktseed$, to the electron
temperature is about $10^{-4}$ (ZLS99; \citealt{gcr00}). For
$\kte\sim 100$ keV this gives $\ktseed\sim 10$ eV which is an
order of magnitude smaller than the disc temperature in BHBs and
closer to that expected from Seyferts. For $\ktseed \sim
300$ eV, the spectra are too hard for the given
reflection fraction. In this model, the spectral slope is an
extremely steep function of the overlap between the corona and
the disc, while the reflection varies very little \citep{bel01}.
Intrinsic dissipation in the disc can make spectra softer
for the given reflection, but then the spectral slope will be
even steeper function of the overlap. Values of
reflection larger than $1$ sometimes observed in Seyferts also
cannot be explained. On the other hand, all the data can be well
described by the ejection model with the correct $\ktseed$
(\citealt{bel99,bel99b}; MBP01).

\begin{figure*}
\centerline{\epsfig{file=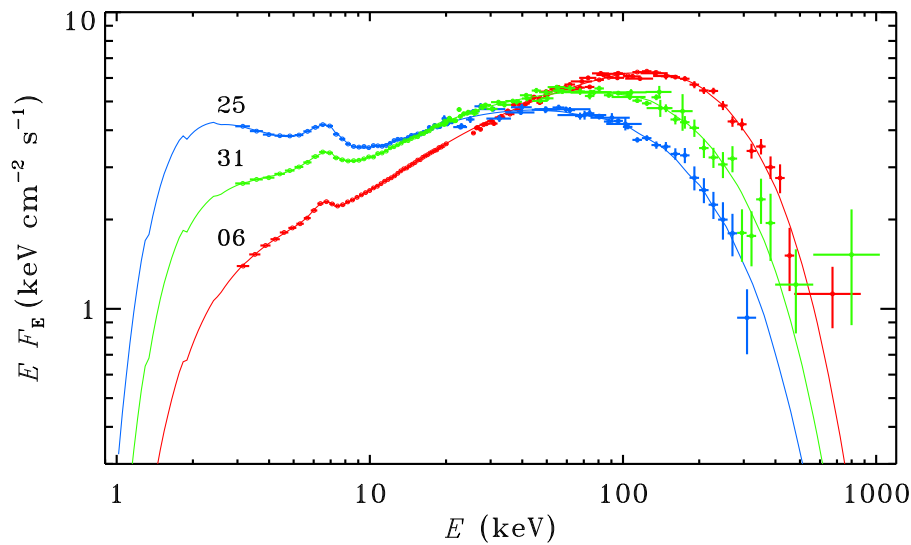,width=12.cm}}
\caption{Broad-band spectra of Cygnus X-1 as observed by \xte\/
and OSSE (with the respective observation number from
Table~\ref{tab:obslog}) and the best-fitting models. Theoretical
curves represent model 1 for observations 6, 25 and 31 (see
Sect. \ref{sec:analysis}). } \label{spectra}
\end{figure*}

The problem is that theoretical models try to reproduce the best-fitting
$\Gamma$ and $\Refl$ which are subject to a number of systematic effects. The
photon index and, especially, the strength of the Compton reflection depend
crucially on the spectral shape of the underlying continuum and description of
the reflection physics (see detailed discussion in \citealt{zdz03}). All the
papers above, where this correlation was studies, assumed the underlying
spectrum to be a power-law. Since the Comptonization spectrum has  a cut-off at
high energies it gives fewer incident photons that are available for reflection
\citep*{w98,per02,mp02}, and thus the fitted $\Refl$ would be larger. Fitting an
exponentially cut-off power-law to the broad-band data (e.g.
\citealt*{matt01,per02}) does not improve the situation, since this model does
not reproduce well the shape of the Comptonization continuum \citep{mp02,zdz03}.
Approximate treatment of ionization could be another source of errors.

As we discussed above, the spectra of BHBs are rather complex having  number of
components, and it is not possible to resolve different spectral components
(e.g. thermal Comptonization and soft excess) in the narrow energy range of an
instrument such as \xte/PCA or \Ginga/LAC. There is a danger that different
components  overlap in that energy band producing effectively a power-law of one
index while in reality the slope of the primary emission could be rather
different. The resulting reflection amplitude  could be also influenced
significantly. Thus, in order to obtain actual $\Gamma$ and $\Refl$ to  be used
in theoretical models, analysis of broad-band data with physical models (such as
Comptonization) is absolutely necessary.

Additional sources of information are the width of the fluorescent Fe K$\alpha$ line at
6.4 keV and the frequencies of the quasi-periodic oscillations that were also
observed  to correlate with the reflection fraction and the spectral slope
\citep[][GCR99]{gcr00}. This seems to be consistent with the variations of the
inner cold disc radius. Not much data exist on the variability of the electron
temperature and Thomson optical depth of the Comptonizing source which can
provide information about the nature of the emitting plasma (electron-proton or
electron-positron). It would be of interest to determine how the
optical depth changes with the bolometric flux, because this can help in
distinguishing the accretion mode the flow is in.

In this paper, we analyze a large set of simultaneous broad-band spectra. Four
observations of Cyg X-1  by \Ginga/LAC and \gro/OSSE from 1991 as well as 38
observations by \xte/PCA, \xte/HEXTE and \gro/OSSE from 1996 -- 1999 are
studied in details.  For the spectral analysis, we use physically motivated
Comptonization models and study correlations between model parameters such as
spectral slope of the primary Comptonization continuum, reflection amplitude,
the width of the Fe line, electron temperature of the hot gas, and its Thomson
optical depth.


\section{OBSERVATIONS AND DATA ANALYSIS}
\label{sec:data}

The observation log is presented in Table
\ref{tab:obslog}. Data reduction for \xte\ has been carried out
using LHEASOFT 5.3.1 software; PCA responses were generated using
pcarsp v.\ 10.1 and HEXTE responses were used from 2000 May 26.
In PCA data reduction, all 5 PCUs were taken into account,
when possible. If not all PCUs were turned on, we use PCUs 0, 2
and 3. Judging from the Crab data, these two PCU configurations
produce similar spectral slopes. Systematic errors of 0.5\% were
added in quadrature to the PCA data. \gro/OSSE spectra were
prepared by adding per-orbit data (average exposure 2--5 ksec),
with total exposure up to 12 hours and contemporary PCA
observation in the middle of the period. The systematic error in
OSSE spectra varies from 3\% at 50 keV to 0.3\% at 300 keV.
Stability of OSSE spectra was checked using hardness ratios
(154--282/52--154 keV) for per-orbit spectra being added. If
orbital spectra within 12 hours were apparently different, we
lowered the total integration time to include only similar data.

We used  PCA data from 3 to 20 keV, HEXTE data from 20-25 to 200
keV and OSSE data from 50 to 1000 keV.
In addition, four \Ginga/LAC and OSSE
simultaneous observations from 1991, previously studied by
\citet{gier97}, were also analyzed. The \Ginga/LAC data are
available from 1.7 keV, but we decided to use exactly the same
energy interval as covered by the PCA.
 For the spectral analysis, we use XSPEC 11.3.1k \citep{arn96}.

The spectra may be separated into two groups based on the difference of their
photon index (see Fig.~\ref{spectra}).
Spectra from 1991 and 1997, represented  by observation 6,
have $\Gamma \sim 1.6$ (usual for the hard state) and
ones from 1999 (observations 25 and 31) have $\Gamma \sim 1.8$--2.2,
and hereafter we call the corresponding state flat
(because the spectrum is nearly flat in the $E F_\mathrm{E}$ plot).
The spectra from 1996 and 1998 are close to the hardest ones from 1999.
We do not consider here the broad-band soft state spectra,
for which the similar analysis has been performed  by G99 and \citet{fro01}.

\begin{table*}
\begin{minipage}{200mm}\caption{Observation log}
\fontsize{8pt}{4mm} \selectfont
\begin{tabular}{|llcclclcl|}
\hline
No & \xte\ & \Ginga\ or PCA        & HEXTE & Date   &  Time [UT]   & \gro  & OSSE         &   Time [UT]     \\
   & ObsID & exposure$^a$ [s] &  Exposure$^b$ [s] &  &         &  VP   & exposure [s] &   \\
 \hline  \multicolumn{9}{|c|}{1991 hard state}\\
G1&& 2304 & & June 6  & 00:18--02:10 & 02 & 4121 & 00:03--02:11 \\
G2&& 888  & & June 6  & 04:43--06:29 & 02 & 4040 & 04:29--06:51 \\
G3&& 2828 & & June 6  & 11:03--14:25 & 02 & 5975 & 10:43--14:32 \\
G4&& 1272 & & June 6  & 20:22--20:44 & 02 & 1629 & 20:02--20:33 \\
\hline
\multicolumn{9}{|c|}{1996 flat state  }\\
1  & 10238-01-08-000 & 14592  & 5083 & March 26 & 10:12--17:36 & 516.5   &17538 &   07:22--20:01 \\
2  & 10238-01-07-000 & 8446  & 2922 & March 27 & 23:06--05:20 & 516.5   &23018 &   20:20--07:47 \\
3  & 10238-01-06-00  & 11455 & 3713 & March 29 & 11:43--17:33 & 516.5   &21868 &   08:24--21:06 \\
4  & 10238-01-05-000 & 10397 & 3524& March 30 & 19:54--01:58$^c$ & 516.5&24153&   16:51--04:12$^c$ \\
\hline
\multicolumn{9}{|c|}{1997 hard state}\\
5    & 10239-01-01-00 & 9095  & --& Feb 2  & 20:13--02:03 & 612.5 &47864 & 16:53--04:52$^c$ \\
6    & 10238-01-03-00 & 6441 & 1938  & Feb 3  & 19:30--22:06 & 612.5 &44530 & 14:38--02:36$^c$ \\
7    & 30158-01-01-00 & 1175 & 809 & Dec 10 & 07:08--08:30 & 705   &7552  & 02:55--12:46 \\
8    & 30158-01-02-00 & 2012 & 823 & Dec 11 & 07:06--08:45 & 705   &12004 & 04:11--14:05\\
9    & 30158-01-03-00 & 2027 & 706 & Dec 14 & 08:48--10:20 & 705   &11599 & 03:09--11:35\\
10   & 30158-01-05-00 & 2614 & 901 & Dec 15 & 05:26--07:09 & 705   &16721 & 23:31--11:18$^c$\\
11   & 30158-01-06-00 & 3210 & 941 & Dec 17 & 00:40--02:05 & 706   &13017 & 20:13--07:32$^d$\\
12   & 30157-01-02-00 & 2309 & 784 & Dec 18 & 07:07--08:16 & 706   &9924  & 01:49--08:46\\
13   & 30158-01-07-00 & 2275 & 766 & Dec 20 & 07:11--08:29 & 706   &11847 & 02:46--12:52\\
14   & 30158-01-08-00 & 2581 & 878 & Dec 21 & 05:28--07:05 & 706   &14924 & 00:51--12:31$^c$\\
15   & 30157-01-03-00 & 2846 & 860 & Dec 24 & 21:24--23:03 & 707   &10127 & 21:36--03:18$^c$\\
16   & 30161-01-01-000& 13244& 4136 & Dec 28 & 13:56--21:03 & 707   &13717 & 11:33--00:24$^c$\\
17   & 30158-01-12-00 & 2836 & 916 & Dec 30 & 03:52--05:00 & 707   & 8979 & 23:49--07:57$^d$ \\
\hline
\multicolumn{9}{|c|}{1998 flat state }\\
18 & 30155-01-01-020  & 10027 & 3897 & Dec 23 & 00:07--05:58 & 804    & 2957  &    21:52--09:00$^d$\\
19 & 30155-01-02-00   & 9625  & 3399 & Dec 28 & 01:40--07:09 & 804    & 5124  &    23:14--10:29$^d$\\
20 & 30161-01-03-01   & 9863  & 3371 & Dec 28 & 13:08--18:37 & 804    & 5289  &   11:50--21:31\\
21 & 40100-01-04-00   & 8472  & --& Dec 29 & 01:37--05:57 & 804    & 5476  &   22:51--10:08$^d$\\
22 & 40100-01-05-00   & 8476  & --& Dec 30 & 01:38--05:57 & 804    & 5288  &   22:30--09:46$^d$\\
23 & 40100-01-06-00   & 7245  & --& Dec 31 & 03:12--07:11 & 804    & 5552  &   23:44--11:00$^d$\\
\hline
\multicolumn{9}{|c|}{1999 flat state }\\
24   & 40101-01-09-00 & 2405 & 665 & Oct 5  & 18:39--19:45 & 831.5   &29605 &   15:04--01:12$^c$\\
25   & 40101-01-11-00 & 731  & 165 & Oct 6  & 19:22--20:03 & 831.5   &25960 &   14:39--00:50$^c$ \\
26   & 40101-01-12-00 & 865  & --& Oct 7  & 07:10--07:40 & 831.5   &28389 &   03:08--12:55\\
27   & 40101-01-15-00 & 741  & 218 & Oct 8  & 08:07--08:48 & 831.5   &31014 &   04:20--14:35\\
28   & 40101-01-16-00 & 757  & 215 & Oct 9  & 09:41--10:22 & 831.5   &33770 &   03:55--15:48\\
29   & 40099-01-20-01 & 1228 & 354 & Oct 12 & 17:33--19:56 & 831.5   &37186 &   12:11--01:23$^c$\\
30   & 40100-01-11-01 & 4300 & --& Oct 28 & 10:46--15:12 & 832     &19559 &   06:16--17:46\\
31   & 40099-01-22-00 & 1444 & 520 & Nov 8  & 14:44--15:25 & 832     &17504 &   09:52--21:28\\
32   & 40099-01-23-01 & 3084 & 1453 & Nov 23 & 15:26--17:32 & 834.5   &8734  &   15:14--21:54\\
33   & 40100-01-13-01 & 770 & --& Nov 24 & 20:08--20:59 & 834.5   &19652 &   14:51--02:13$^c$\\
34   & 40100-01-14-02 & 479 & --& Nov 25 & 20:05--20:56 & 834.5   &16683 &   14:24--01:47$^c$\\
35   & 40100-01-15-03 & 1729 & --& Nov 26 & 21:40--22:32 & 834.5   &24372 &   17:11--04:33$^c$\\
36   & 40100-01-16-02 & 1448 & --& Nov 27 & 19:59--20:50 & 834.5   &19063 &   15:13--02:34$^c$\\
37   & 40100-01-17-03 & 1902 & --& Nov 28 & 21:32--22:27 & 834.5   &10309 &   19:19--00:28$^c$\\
38   & 40100-01-18-03 & 1961 & --& Nov 29 & 21:29--22:25 & 834.5   &19999 &   16:03--03:19$^c$\\
\hline
\multicolumn{9}{l}{$^a$ The deadtime-corrected exposure.}\\
\multicolumn{9}{l}{$^b$ HEXTE cluster 0 exposure. Symbol '--' means that no HEXTE data are available.} \\
\multicolumn{9}{l}{$^c$ The observation finished on the following day.}\\
\multicolumn{9}{l}{$^d$ The observation started on the previous day.}\\
\\
\end{tabular}
 \label{tab:obslog}
\end{minipage}
\end{table*}

\begin{figure*}
\centerline{\epsfig{file=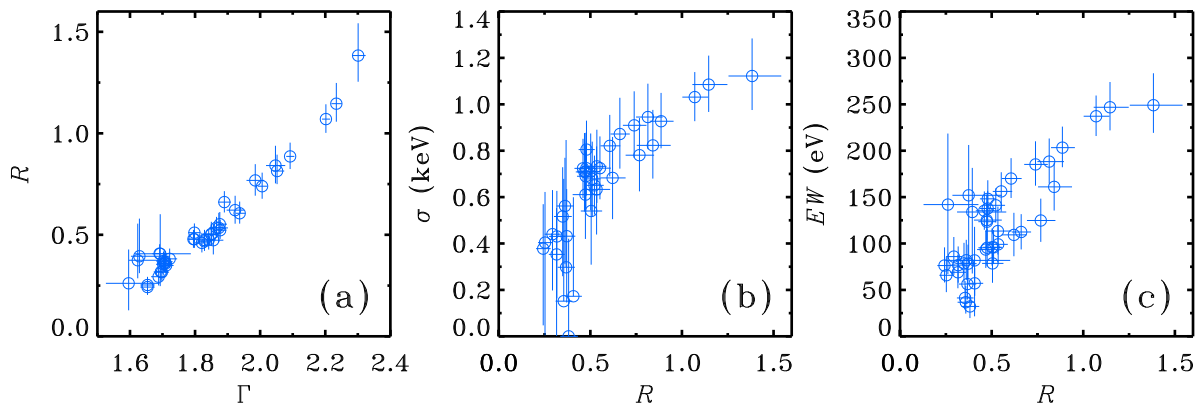,width=16.cm}}
\caption{Correlations between parameters for the model 0 ({\sf pexrav}).
(a) The reflection amplitude, $\Refl$, vs.\ the photon spectral index, $\Gamma$.
(b) The relativistic smearing Gaussian width $\sigma$ vs.\ reflection $\Refl$.
(c) The Fe K$\alpha$ 6.4 keV line equivalent  width  $EW$ vs. $\Refl$.
} \label{fig:pexravrel}
\end{figure*}

\section{SPECTRAL ANALYSIS}
\label{sec:analysis}

In order to describe the observed broad-band spectra we use the following models:

(0) Power-law (without or with an exponential cutoff) and
  Compton reflection ({\sf pexrav} model, \citealt*{mz95});

(1) two thermal Comptonization models and reflection, with the
soft excess also modelled
 by thermal Comptonization;

(2) thermal Comptonization with reflection plus a nonthermal
Comptonization  corresponding to the soft state spectrum.

We use the {\sf eqpair} code (\citealt*{cop99}; G99)  for modeling
thermal and nonthermal Comptonization. All models include also
a Gaussian line  at 6.4 keV and
interstellar absorption with column density $\nh$ which we find
often larger than the value of $0.6\times 10^{22}$ cm$^{-2}$ (derived from the
reddening towards the companion star, see \citealt{bc95}).  To avoid unreasonably
low values of this parameter, its low limit was set to $0.5\times 10^{22}$ cm$^{-2}$.
The presented uncertainties are given at a 90 per cent confidence
level for a single parameter ($\Delta\chi^2=2.71$).
Fluxes, unless stated otherwise, correspond to the
range covering all the model emission.

\subsection{The power-law and reflection model in the 3--20 keV data}
\label{pexravm}

The spectra of Cyg X-1 clearly show correlations between reflection amplitude
and the spectral index  (GCR99, ZLS99).  Detailed analysis confirms
\citep{zdz03} that the extend of correlation is much larger than typical errors
in the best-fitting parameters. However, since the PCA spectrum
falls in a quite narrow energy interval and it is difficult to
distinguish between various spectral components that may form an
``effective" power-law. Therefore, it is not certain that the
values for $\Gamma$ and $\Refl$ obtained from the simple power-law/reflection
fits indeed correspond to the actual physical situation.

Still, in order to compare our results with those of previous
analysis, we have performed fits similar to those presented in
ZLS99 and GCR99. We later compare them to the results obtained
with more physical models (see Sect.~\ref{sec:old_vs_new}). We use
the XSPEC model {\sf phabs*(pexrav+gaussian)} (model 0), i.e., a
power law with the photon index, $\Gamma$, and Compton reflection
with the relative strength, $\Refl$ \citep{mz95}, accompanied by a
Gaussian fluorescence Fe K$\alpha$ line (characterized by the
relativistic smearing width $\sigma$ and the equivalent width
$EW$), all absorbed by interstellar material of column density
$\nh$. Hereafter, we assume the disc inclination of $i=50\degr$
and neutral reflector.

The fit results are presented in Table~\ref{tab:pexrav}. Only the
low-energy 3--20 keV (\xte/PCA and \Ginga/LAC) data were fitted, and
therefore we did not apply any high-energy cutoff to the power
law.
We find that the correlations between spectral parameters (see
Fig.~\ref{fig:pexravrel}) are similar to GCR99 results.

As shown in Sect. \ref{sec:obs1999} below, there is a likely
overlap of different spectral components in the PCA energy range.
Therefore, the model 0 cannot
represent a good approach for physical interpretations for the
spectra.  Taking into account also the OSSE data and including an
exponential cutoff in the model, we can model the joint data only
very roughly, with $\chi^2$/dof $\sim 2$. This is likely to be due
to the the shape of the exponential cutoff (which is assumed in
{\sf pexrav}) being substantially different from the shape of the
cutoff of thermal Comptonization (see e.g. \citealt{zdz03}). This
provides an argument against utilizing simple phenomenological
models in the analysis of BHB spectra.

\begin{table*}
\begin{minipage}{126mm}\caption{The best-fit parameters for model 0
(fitted to the low-energy \Ginga~and \xte/PCA data only).
}
\fontsize{9pt}{4mm} \selectfont
\begin{center}
\begin{tabular}{ccccccc}
\hline
No. &$\nh$ [$10^{22}$ cm$^{-2}$]&$\Gamma$ & $\Refl$ & $\sigma$ [keV]  & $EW$ [eV] & $\chi^2/{\rm dof}$ \\
\hline
\multicolumn{7}{|c|}{1991}\\
G1 &$ 0.5_{-0}^{+ 1.2 }   $&$1.63_{-  0.03}^{+ 0.09} $&$0.39_{-  0.08}^{+ 0.19} $&$0.46_{- 0.46}^{+ 0.81}   $&$134_{-  66}^{+ 47}$& 3/11 \\
G2 &$ 0.5_{-0}^{+ 1.2 }   $&$1.62_{-  0.03}^{+ 0.09} $&$0.37_{-  0.09}^{+ 0.18} $&$0.58_{- 0.58}^{+ 0.93}   $&$152_{-  73}^{+ 54}$& 4/11 \\
G3 &$ 1.3_{-0.8}^{+ 1.3 } $&$1.60_{-  0.07}^{+ 0.10} $&$0.26_{-  0.13}^{+ 0.17} $&$0.60_{- 0.60}^{+ 0.95}   $&$142_{-  75}^{+ 77}$& 3/11 \\
G4 &$ 1.4_{-0.9}^{+ 1.2 } $&$1.69_{-  0.08}^{+ 0.09} $&$0.41_{-  0.16}^{+ 0.19} $&$0.42_{- 0.42}^{+ 0.97}   $&$82_{-  60}^{+ 36} $& 5/12 \\
\multicolumn{7}{|c|}{1996}\\
01 &$ 1.2\pm0.2           $&$1.80 \pm0.02            $&$0.51\pm  0.05           $&$0.68    \pm         0.16 $&$96 \pm 19         $& 24/43\\
02 &$ 0.6\pm0.2           $&$1.79 \pm0.02            $&$0.48\pm  0.05           $&$0.80    \pm         0.12 $&$148\pm 20	     $& 20/43\\
03 &$ 0.7\pm0.2           $&$1.80 \pm0.02            $&$0.48\pm  0.04           $&$0.72    \pm         0.12 $&$138\pm 19	     $& 20/43\\
04 &$ 1.2\pm0.2           $&$1.89 \pm0.02            $&$0.66\pm  0.05           $&$0.87    \pm         0.16 $&$112\pm 19	     $& 24/43\\
\multicolumn{7}{|c|}{1997}\\
05 &$ 1.3\pm0.3           $&$1.65 \pm0.02            $&$0.25\pm  0.04           $&$0.40_{-0.40}^{+ 0.22} $&$66 \pm 20	     $& 19/39\\
06 &$ 1.2\pm0.3           $&$1.65 \pm0.02            $&$0.24\pm  0.04           $&$0.38_{-0.33}^{+ 0.19} $&$76 \pm 20	     $& 27/39\\
07 &$ 1.0\pm0.3           $&$1.71 \pm0.02            $&$0.35\pm  0.05           $&$0.52_{-0.26}^{+ 0.21} $&$79 \pm 22	     $& 16/39\\
08 &$ 1.0\pm0.3           $&$1.69 \pm0.02            $&$0.29\pm  0.04           $&$0.44_{-0.24}^{+ 0.19} $&$86 \pm 21	     $& 35/39\\
09 &$ 1.2\pm0.3           $&$1.71 \pm0.02            $&$0.36\pm  0.05           $&$0.56_{-0.25}^{+ 0.21} $&$82 \pm 22	     $& 26/39\\
10 &$ 1.2\pm0.2           $&$1.69 \pm0.02            $&$0.32\pm  0.05           $&$0.43_{-0.27}^{+ 0.20} $&$77 \pm 20	     $& 22/39\\
11 &$ 1.2\pm0.2           $&$1.70 \pm0.02            $&$0.32\pm  0.04           $&$0.35_{-0.35}^{+ 0.20} $&$69 \pm 19	     $& 15/39\\
12 &$ 3.1\pm0.2           $&$1.72 \pm0.02            $&$0.38\pm  0.05           $&$0_{-0}^{+ 0.42  }$&$32 \pm 12	     $& 23/39\\
13 &$ 2.1\pm0.3           $&$1.71 \pm0.02            $&$0.37\pm  0.05           $&$0.30_{-0.30}^{+ 0.55  } $&$ 56\pm19	     $& 21/39\\
14 &$ 1.1\pm0.3           $&$1.70 \pm0.02            $&$0.37\pm  0.05           $&$0.43_{-0.29}^{+ 0.20}   $&$78 \pm21	     $& 21/39\\
15 &$ 2.1\pm0.3           $&$1.70 \pm0.02            $&$0.35\pm  0.05           $&$0.15_{-0.15}^{+ 0.53}    $&$41 \pm18	     $& 17/39\\
16 &$ 2.1\pm0.2           $&$1.69 \pm0.02            $&$0.41\pm  0.05           $&$0.17_{-0.17}^{+ 0.43}    $&$57 \pm16	     $& 22/39\\
17 &$ 2.7\pm0.3           $&$1.71 \pm0.02            $&$0.36\pm  0.05           $&$0.03_{-0.03}^{+ 2.79}    $&$37 \pm17	     $& 19/39\\
\multicolumn{7}{|c|}{1998}\\
18 &$ 1.2\pm0.2           $&$1.87 \pm0.02            $&$0.54\pm  0.05           $&$0.73    \pm    0.16   $&$99 \pm 20	     $& 19/39\\
19 &$ 1.2\pm0.2           $&$1.84 \pm0.02            $&$0.48\pm  0.05           $&$0.71    \pm    0.17   $&$95 \pm 20	     $& 19/39\\
20 &$ 1.2\pm0.2           $&$1.85 \pm0.02            $&$0.50\pm  0.05           $&$0.71    \pm    0.17   $&$96 \pm 20	     $& 17/39\\
21 &$ 1.1\pm0.2           $&$1.83 \pm0.02            $&$0.47\pm  0.05           $&$0.71    \pm    0.17   $&$93 \pm 20	     $& 14/39\\
22 &$ 0.7\pm0.2           $&$1.83 \pm0.02            $&$0.47\pm  0.05           $&$0.69    \pm    0.14   $&$123\pm 21	     $& 20/39\\
23 &$ 0.7\pm0.2           $&$1.82 \pm0.02            $&$0.46\pm  0.05           $&$0.72    \pm    0.13   $&$135\pm 21	     $& 17/39\\
\multicolumn{7}{|c|}{1999}\\								      
24 &$ 0.5_{-0}^{+ 0.1 }   $&$2.09 \pm0.01            $&$0.89\pm  0.07           $&$0.93    \pm    0.12   $&$203\pm 23	     $& 28/33\\
25 &$ 0.5_{-0}^{+ 0.1   } $&$2.30 \pm0.02            $&$1.38\pm  0.16           $&$1.12    \pm    0.16   $&$249\pm 34	     $& 48/33\\
26 &$ 0.5_{-0}^{+ 0.1   } $&$2.23 \pm0.01            $&$1.15\pm  0.10           $&$1.09    \pm    0.13   $&$247\pm 27	     $& 43/33\\
27 &$ 0.5_{-0}^{+ 0.2  }  $&$2.05 \pm0.01            $&$0.81\pm  0.08           $&$0.95    \pm    0.14   $&$188\pm 25	     $& 14/33\\
28 &$ 0.7\pm0.3           $&$2.05 \pm0.03            $&$0.84\pm  0.10           $&$0.82    \pm    0.15   $&$161\pm 26	     $& 32/33\\
29 &$ 0.5_{-  0}^{+ 0.1 } $&$2.01 \pm0.01            $&$0.74\pm  0.07           $&$0.91    \pm    0.15   $&$185\pm 25	     $& 22/33\\
30 &$ 0.5_{-  0}^{+ 0.1 } $&$2.20 \pm0.01            $&$1.07\pm  0.07           $&$1.03    \pm    0.11   $&$237\pm 22	     $& 27/33\\
31 &$ 0.5_{-  0}^{+ 0.2 } $&$1.94 \pm0.02            $&$0.61\pm  0.06           $&$0.82    \pm    0.13   $&$170\pm 22	     $& 25/33\\
32 &$ 1.1\pm0.3           $&$1.85 \pm0.03            $&$0.50\pm  0.06           $&$0.54    \pm    0.23   $&$79 \pm 22	     $& 20/33\\
33 &$ 0.7_{-0.2}^{+ 0.3}  $&$1.87 \pm0.03            $&$0.53\pm  0.07           $&$0.63    \pm    0.19   $&$113\pm 26	     $& 30/33\\
34 &$ 0.6_{-0.1}^{+ 0.3}  $&$1.86 \pm0.03            $&$0.47\pm  0.08           $&$0.61    \pm    0.19   $&$125\pm 28	     $& 14/33\\
35 &$ 0.5_{-0}^{+ 0.2 }   $&$1.88 \pm0.02            $&$0.52\pm  0.05           $&$0.65    \pm    0.13   $&$141\pm 18	     $& 25/33\\
36 &$ 0.5_{-0}^{+ 0.2 }   $&$1.88 \pm0.02            $&$0.55\pm  0.06           $&$0.72    \pm    0.14   $&$156\pm 22	     $& 25/33\\
37 &$ 1.2\pm0.3           $&$1.92 \pm0.03            $&$0.62\pm  0.07           $&$0.68    \pm    0.18   $&$109\pm 24	     $& 27/33\\
38 &$ 1.2\pm0.3           $&$1.98 \pm0.03            $&$0.77\pm  0.08           $&$0.78    \pm    0.16   $&$125\pm 24	     $& 39/33\\							      
\hline											      
\end{tabular}
\end{center}
\label{tab:pexrav}
\end{minipage}
\end{table*}

\subsection{Comptonization model and broad-band spectra}
\label{sec:obs97} \label{sec:obs1999}

 The hard state spectra are well described
by thermal Comptonization \citep{gier97,p98,fro01},
with a weak soft excess.
To describe Comptonization we use  the XSPEC model {\sf eqpair }
(see \citealt*{cop99}; G99). The spectrum
of seed photons is from a pseudo-Newtonian accretion disc, see
G99. Parameters of emission are expressed through the compactness,
\be \ell={L\sigmat\over {\cal R}\me c^3}, \ee where $L$ is the
source luminosity, ${\cal R}$ is the radius of the emitting
spherical cloud, and $\sigma_\mathrm{T}$ is the Thomson cross
section. We consider here thermal plasma. The model is
characterized by the following parameters: $\ls$ -- the
compactness of soft seed photons (assumed here to be 1); $\lhls$
-- the ratio of the dimensionless energy dissipation rate in a hot
cloud to $\ls$; $\ktbb$ -- the maximal colour temperature of the
disc (at 9.5 gravitational radii, $\rg=GM/c^2$, or at the inner
radius of the disc, if it is cut off at larger value); $\taub$ --
the Thomson optical depth corresponding to the ions; $\Refl$ --
the reflection amplitude. Since there are large  ($\sim$25\%
at $1\sigma$)
 errors on $\tin$, we decided to fix it at
the common value of 200 eV. The model computes the coronal
temperature, $\kte$, and the total optical depth, $\taut$, from
the background electrons and produced $e^\pm$ pairs
self-consistently from the energy and pair balance. For the
assumed compactness, pair production is negligible for all
considered spectra, and thus the resulting total optical depth
$\taut=\taub$. In the fits, we assume the inner disc radius of
$6\rg$ (this parameter has only effect on the relativistic
smearing of the reflected component, that could be hardly resolved
with present energy resolution).

 The normalization of {\sf eqpair}, corresponding to the
disc component, is $f_cM^2\cos i/(D^2\betac^4)$, where $f_c$ is
the covering factor, $M$ is the black hole mass in units of the
solar mass, $D$ is the distance to the source in units of kpc, and
$\betac$ is the ratio of the colour temperature to the effective
one.

At low energies the effect of the interstellar absorption
is clearly visible.   The addition of a additional  soft component
leads to further improvement of the fits. Such soft excesses were
observed in both the hard state by \citet{ds01} and in a flat
state (similar to that analyzed here) by \citet{fro01} in \sax\/
data. This component, that nature we address below, however, is
relatively weak in case of 1991 and 1997 (observations G1--G4,
05--17) hard state data. The spectra observed in 1996, 1998 and
1999 (observations 1--4, 18--38) appear similar in the overall
shape to those of hard state, but are significantly softer (see
Fig.~\ref{fig:pexravrel}).  The soft excess in the flat state is
stronger.

We stress that the requirement of an additional soft excess is
implied only by the joint PCA/HEXTE/OSSE data, since the PCA data
cover a too narrow energy range. Even if the actual spectrum in
the PCA 3--20 keV band is not a power-law but is e.g. concave, a
good fit with a power law plus reflection (model 0) can be still
achieved. However, the real strength of Compton reflection can be
significantly different.

Since the data require an additional component only in a
relatively narrow, $\sim$3--10 keV, range, the parameters of the
soft excess cannot be constrained accurately. Below we consider a
number of physically realistic scenarios of its nature. In each of
the considered models, we restrict the parameters controlling its
spectral shape to values that makes its flux significant only at
low energies, and we fit only its normalization. These fits allow
us to completely describe the broad-band spectra and to constrain
the parameters of main continuum.


\begin{table*}
\begin{minipage}{180mm}
\fontsize{8pt}{4mm} \selectfont \caption{The best-fit parameters
for model 1.}
\begin{center}
\begin{tabular}{ccccccccccccc}
\hline Obs. & $\nh^a$ &$\lhls$                      & $\taut$                      & $\Refl$         & $\sigma$ & $EW$ &$N^b$ &$F_\mathrm{tot}^c$ & $F_\mathrm{add}^d$  & $\kte^e$ &  $\Gamma^f$ & $\chi^2/{\rm dof}$ \\
 \hline\hline
G1 & $1.1_{- 0.6}^{+ 0.8}$ &  $13.8 _{-0.7}^{+ 0.9 } $   & $1.39 _{-0.08}^{+ 0.07}$ &  $0.38_{- 0.03}^{+ 0.04}$  &  $0.67_{- 0.34}^{+ 0.32}$  &  $189_{- 51}^{+ 57}$&   $1.36 _{-0.06}^{+ 0.05}$    & 5.32  &  0.11  &  90       &  1.62   & 49/73  \\
G2 & $1.3\pm 0.8         $ &  $15.6 _{-0.8}^{+ 1.3 } $   & $1.58 _{-0.05}^{+ 0.12}$ &  $0.35_{- 0.05}^{+ 0.04}$  &  $0.87_{- 0.31}^{+ 0.36}$  &  $235_{- 63}^{+ 75}$&   $1.63 _{-0.09}^{+ 0.06}$    & 7.24  &  0.24  &  82       &  1.59   & 58/74  \\
G3 & $1.4\pm 0.2         $ &  $14.0 _{-1.6}^{+ 0.9 } $   & $1.34 _{-0.20}^{+ 0.08}$ &  $0.26_{- 0.04}^{+ 0.06}$  &  $0.60_{- 0.60}^{+0.92} $  &  $151_{- 47}^{+ 54}$&   $1.13 _{-0.04}^{+ 0.10}$    & 4.18  &  0     &  94       &  1.62   & 53/73   \\
G4 & $2.0_{- 0.9}^{+ 0.7}$ &  $13.4\pm 1.8           $   & $1.36 _{-0.30}^{+ 0.29}$ &  $0.26_{- 0.08}^{+ 0.09}$  &  $0.63_{- 0.31}^{+ 0.41}$  &  $132_{- 48}^{+ 57}$&   $1.04 _{-0.11}^{+ 0.14}$    & 3.83  &  0.14  &  90       &  1.63  & 53/73   \\
01 & $2.6\pm 0.2         $ &  $10.3 _{-0.1}^{+ 0.4 } $   & $1.31 _{-0.03}^{+ 0.06}$ &  $0.27_{- 0.01}^{+ 0.02}$  &  $1.00_{- 0.13}^{+ 0.16}$  &  $173_{- 20}^{+ 26}$&   $1.99 _{-0.09}^{+ 0.03}$    & 7.32  &  1.26  &  86       &  1.68   & 501/478 \\
02 & $2.1\pm 0.3         $ &  $9.76_{-0.24}^{+ 0.18} $   & $1.32 _{-0.03}^{+ 0.02}$ &  $0.29_{- 0.02}^{+ 0.01}$  &  $1.11_{- 0.13}^{+ 0.12}$  &  $240_{- 27}^{+ 25}$ &  $2.50 _{-0.04}^{+ 0.06}$    & 8.83  &  1.56  &  83       &  1.69   & 531/493 \\
03 & $2.2_{- 0.2}^{+ 0.3}$ &  $10.2 _{-0.2}^{+ 0.1 } $   & $1.47 _{-0.03}^{+ 0.05}$ &  $0.26\pm 0.02          $  &  $1.02_{- 0.13}^{+ 0.14}$  &  $227_{- 29}^{+ 28}$&   $2.18\pm 0.03          $    & 8.04  &  1.49  &  75       &  1.68   & 542/506 \\
04 & $2.9\pm 0.2         $ &  $8.77_{-0.14}^{+ 0.18} $   & $1.38 _{-0.02}^{+ 0.03}$ &  $0.31\pm 0.01          $  &  $1.12\pm 0.11          $  &  $215_{- 19}^{+ 23}$ &  $2.85 _{-0.07}^{+ 0.05}$    & 10.0  &  2.44  &  77       &  1.71   & 529/493 \\
05 & $1.6_{- 0.3}^{+ 0.4}$ &  $11.8 _{-0.4}^{+ 1.2 } $   & $1.09 _{-0.03}^{+ 0.16}$ &  $0.22\pm 0.02          $  &  $0.48_{- 0.23}^{+ 0.22}$  &  $77 _{- 16}^{+ 18}$&   $1.30 _{-0.09}^{+ 0.03}$    & 4.46  &  0.08  &  110      &  1.65  & 61/73  \\
06 & $1.8\pm 0.4         $ &  $12.8 _{-0.5}^{+ 0.3 } $   & $1.29 _{-0.08}^{+ 0.03}$ &  $0.19_{- 0.01}^{+ 0.02}$  &  $0.54_{- 0.21}^{+ 0.19}$  &  $96 \pm 19        $&   $1.05 _{-0.02}^{+ 0.04}$    & 3.99  &  0.19  &  95       &  1.64   & 374/402 \\
07 & $1.9_{- 0.4}^{+ 0.3}$ &  $12.0\pm 0.4           $   & $1.24 _{-0.07}^{+ 0.13}$ &  $0.22\pm 0.02          $  &  $0.75_{- 0.23}^{+ 0.19}$  &  $121_{- 27}^{+ 24}$&   $1.54 _{-0.06}^{+ 0.04}$    & 5.80  &  0.52  &  96       &  1.65   & 367/415\\
08 & $2.0_{- 0.4}^{+ 0.3}$ &  $11.9 _{-0.4}^{+ 0.3 } $   & $1.41 _{-0.04}^{+ 0.03}$ &  $0.20_{- 0.01}^{+ 0.02}$  &  $0.68_{- 0.21}^{+ 0.19}$  &  $118_{- 22}^{+ 21}$&   $1.47 _{-0.02}^{+ 0.04}$    & 5.43  &  0.46  &  84       &  1.65   & 404/415 \\
09 & $2.3\pm 0.3         $ &  $12.0 _{-0.2}^{+ 0.4 } $   & $1.41 _{-0.06}^{+ 0.03}$ &  $0.23\pm 0.02          $  &  $0.85_{- 0.18}^{+ 0.20}$  &  $131_{- 21}^{+ 26}$&   $1.56 _{-0.04}^{+ 0.03}$    & 5.98  &  0.62  &  84       &  1.65   & 350/415 \\
10 & $1.6_{- 0.4}^{+ 0.5}$ &  $11.3 _{-0.4}^{+ 0.3 } $   & $1.18 _{-0.05}^{+ 0.08}$ &  $0.24_{- 0.01}^{+ 0.02}$  &  $0.56_{- 0.18}^{+ 0.19}$  &  $102_{- 18}^{+ 21}$&   $1.43 _{-0.04}^{+ 0.02}$    & 4.92  &  0.25  &  99       &  1.66   & 301/334 \\
11 & $4.1\pm 0.3         $ &  $11.6 _{-0.5}^{+ 0.4 } $   & $1.19 _{-0.03}^{+ 0.06}$ &  $0.23\pm 0.02          $  &  $0.41_{- 0.41}^{+0.74}    $  &  $54 _{- 17}^{+ 18}$&   $1.53 \pm0.04          $    & 5.72  &  0.58  &  99       &  1.65   & 193/236 \\
12 & $4.2\pm 0.3         $ &  $11.8 _{-0.5}^{+ 0.6 } $   & $1.20 _{-0.03}^{+ 0.10}$ &  $0.23_{- 0.01}^{+ 0.02}$  &  $0.43_{- 0.41}^{+0.78}    $  &  $55 _{- 17}^{+ 20}$&   $1.52 _{-0.06}^{+ 0.04}$    & 5.77  &  0.61  &  99       &  1.65   & 202/238 \\
13 & $2.5_{- 0.8}^{+ 1.0}$ &  $12.8\pm 0.4           $   & $1.45 _{-0.03}^{+ 0.05}$ &  $0.22\pm 0.02          $  &  $0.58_{- 0.18}^{+ 0.24}$  &  $106_{- 20}^{+ 19}$&   $1.45 _{-0.03}^{+ 0.04}$    & 5.74  &  0.47  &  83       &  1.63   & 369/411 \\
14 & $2.2\pm 0.3         $ &  $12.4 _{-0.3}^{+ 0.4 } $   & $1.34\pm 0.03          $ &  $0.24\pm 0.02          $  &  $0.74_{- 0.17}^{+ 0.19}$  &  $126_{- 19}^{+ 23}$&   $1.58 _{-0.04}^{+ 0.03}$    & 6.25  &  0.63  &  90       &  1.64   & 397/415 \\
15 & $3.2\pm 0.3         $ &  $13.6\pm 0.5           $   & $1.26 _{-0.04}^{+ 0.11}$ &  $0.19\pm 0.02          $  &  $0.60_{- 0.30}^{+ 0.35}$  &  $75 _{- 19}^{+ 27}$&   $1.29 _{-0.04}^{+ 0.03}$    & 5.51  &  0.57  &  99       &  1.62   & 201/244 \\
16 & $3.4\pm 0.3         $ &  $13.7\pm 0.2           $   & $1.61 \pm0.04          $ &  $0.24\pm 0.01          $  &  $0.55_{- 0.16}^{+ 0.19}$  &  $98 _{- 14}^{+ 19}$&   $1.47 \pm0.02          $    & 6.53  &  0.76  &  76       &  1.62   & 446/402 \\
17 & $3.8\pm 0.3         $ &  $12.9 _{-0.4}^{+ 0.6 } $   & $1.33 _{-0.10}^{+ 0.15}$ &  $0.19_{- 0.03}^{+ 0.02}$  &  $0.60_{- 0.38}^{+ 0.29}$  &  $73 \pm 21        $&   $1.25 _{-0.05}^{+ 0.02}$    & 5.12  &  0.55  &  92       &  1.63   & 345/415 \\
18 & $2.5\pm 0.3         $ &  $8.13\pm0.15          $    & $1.23 _{-0.03}^{+ 0.02}$ &  $0.28_{- 0.01}^{+ 0.02}$  &  $0.98_{- 0.11}^{+ 0.13}$  &  $182_{- 21}^{+ 23}$&   $2.32 \pm0.04          $    & 7.09  &  1.41  &  84       &  1.73   & 512/489 \\
19 & $2.7\pm 0.3         $ &  $9.43_{-0.09}^{+ 0.23}$    & $1.61 _{-0.03}^{+ 0.04}$ &  $0.21\pm 0.02          $  &  $1.02_{- 0.12}^{+ 0.15}$  &  $182_{- 21}^{+ 25}$&   $1.67 _{-0.03}^{+ 0.02}$    & 5.87  &  1.29  &  66       &  1.69   & 457/502 \\
20 & $2.6\pm 0.3         $ &  $9.14_{-0.09}^{+ 0.20}$    & $1.45 _{-0.06}^{+ 0.03}$ &  $0.24_{- 0.02}^{+ 0.01}$  &  $1.00_{- 0.12}^{+ 0.15}$  &  $182_{- 20}^{+ 26}$&   $1.79 _{-0.03}^{+ 0.02}$    & 6.11  &  1.29  &  73       &  1.70   & 472/502 \\
21 & $2.6\pm 0.4         $ &  $9.91_{-2.68}^{+ 2.16}$    & $1.36 _{-0.43}^{+ 0.59}$ &  $0.20_{- 0.14}^{+ 0.11}$  &  $0.94_{- 0.14}^{+ 0.18}$  &  $161_{- 23}^{+ 46}$&   $1.83 _{-0.42}^{+ 0.56}$    & 6.52  &  1.28  &  81       &  1.68   & 46/73   \\
22 & $2.1_{- 0.4}^{+ 0.3}$ &  $8.96_{-2.05}^{+ 2.38}$    & $1.19 _{-0.55}^{+ 0.43}$ &  $0.24_{- 0.13}^{+ 0.08}$  &  $0.90\pm 0.12          $  &  $188_{- 19}^{+ 25}$&   $2.13 _{-0.55}^{+ 0.58}$    & 6.86  &  1.21  &  90       &  1.71   & 45/73   \\
23 & $2.0\pm 0.4         $ &  $8.57_{-1.42}^{+ 1.55}$    & $1.29 _{-0.65}^{+ 0.25}$ &  $0.26_{- 0.09}^{+ 0.14}$  &  $0.89_{- 0.11}^{+ 0.12}$  &  $192_{- 20}^{+ 30}$&   $2.14 _{-0.16}^{+ 0.58}$    & 6.57  &  1.10  &  81       &  1.72   & 43/73   \\
24 & $1.6_{- 0.2}^{+ 0.3}$ &  $5.43_{-0.19}^{+ 0.68}$    & $1.08 _{-0.03}^{+ 0.13}$ &  $0.36\pm 0.03          $  &  $0.99_{- 0.08}^{+ 0.09}$  &  $337_{- 26}^{+ 28}$&   $1.85 _{-0.27}^{+ 0.06}$    & 4.71  &  1.46  &  83       &  1.82   & 389/409 \\
25 & $1.5\pm 0.2         $ &  $4.70_{-0.29}^{+ 0.57}$    & $1.04 _{-0.06}^{+ 0.17}$ &  $0.36_{- 0.06}^{+ 0.04}$  &  $0.98_{- 0.07}^{+ 0.08}$  &  $395_{- 30}^{+ 33}$&   $2.46 _{-0.24}^{+ 0.14}$    & 6.51  &  2.71  &  81       &  1.85   & 355/409\\
26 & $1.3\pm 0.3         $ &  $4.08_{-0.49}^{+ 0.92}$    & $0.89 _{-0.14}^{+ 0.25}$ &  $0.40_{- 0.09}^{+ 0.08}$  &  $1.01\pm 0.07          $  &  $382_{- 28}^{+ 28}$&   $2.54 _{-0.42}^{+ 0.29}$    & 5.44  &  1.92  &  90       &  1.89   & 48/67   \\
27 & $1.8\pm 0.3         $ &  $5.95\pm0.30          $    & $1.10 _{-0.06}^{+ 0.05}$ &  $0.33\pm 0.03          $  &  $1.02\pm 0.10          $  &  $305\pm 30        $&   $2.19 _{-0.09}^{+ 0.10}$    & 5.92  &  1.81  &  84       &  1.80   & 386/409 \\
28 & $1.7_{- 0.3}^{+ 0.2}$ &  $5.75_{-0.50}^{+ 0.35}$    & $1.21 _{-0.16}^{+ 0.06}$ &  $0.33_{- 0.04}^{+ 0.06}$  &  $0.98_{- 0.08}^{+ 0.09}$  &  $335_{- 27}^{+ 31}$&   $1.74 \pm0.16          $    & 4.71  &  1.52  &  75       &  1.81   & 389/409 \\
29 & $1.6\pm 0.3         $ &  $6.01_{-0.24}^{+ 0.32}$    & $1.09 _{-0.04}^{+ 0.05}$ &  $0.33\pm 0.03          $  &  $0.99_{- 0.10}^{+ 0.11}$  &  $295_{- 29}^{+ 30}$&   $1.98 _{-0.09}^{+ 0.07}$    & 5.06  &  1.28  &  85       &  1.80   & 394/409 \\
30 & $0.9_{- 0.3}^{+ 0.2}$ &  $2.71_{-0.23}^{+ 1.06}$    & $0.46 _{-0.19}^{+ 0.26}$ &  $0.62_{- 0.05}^{+ 0.08}$  &  $0.97\pm 0.08          $  &  $328_{- 25}^{+ 19}$&   $4.06 _{-0.15}^{+ 0.48}$    & 5.67  &  1.46  &  141\dag  &   1.93  & 52/67   \\
31 & $1.7\pm 0.3         $ &  $6.74_{-0.39}^{+ 0.32}$    & $1.09 _{-0.14}^{+ 0.06}$ &  $0.29_{- 0.03}^{+ 0.04}$  &  $0.93 \pm0.11          $  &  $253_{- 26}^{+ 27}$&   $1.91 _{-0.09}^{+ 0.11}$    & 5.10  &  1.12  &  89       &  1.77   & 390/409 \\
32 & $2.4\pm 0.3         $ &  $9.38_{-0.40}^{+ 0.37}$    & $1.37 _{-0.08}^{+ 0.10}$ &  $0.22_{- 0.02}^{+ 0.03}$  &  $0.83\pm 0.15          $  &  $151_{- 25}^{+ 22}$&   $1.49 \pm0.06          $    & 5.08  &  1.01  &  79       &  1.70   & 393/409 \\
33 & $2.1_{- 0.6}^{+ 0.2}$ &  $7.94_{-1.26}^{+ 0.56}$    & $1.23 _{-0.24}^{+ 0.09}$ &  $0.27_{- 0.02}^{+ 0.07}$  &  $0.83_{- 0.19}^{+ 0.15}$  &  $179_{- 32}^{+ 33}$&   $1.93 _{-0.10}^{+ 0.29}$    & 5.76  &  1.16  &  84       &  1.73   & 62/67   \\
34 & $1.5_{- 0.5}^{+ 0.3}$ &  $7.08_{-0.89}^{+ 0.35}$    & $1.08 _{-0.19}^{+ 0.13}$ &  $0.29_{- 0.04}^{+ 0.07}$  &  $0.74_{- 0.18}^{+ 0.17}$  &  $174\pm 31        $&   $1.83 _{-0.11}^{+ 0.25}$    & 4.64  &  0.67  &  92       &  1.76   & 43/67   \\
35 & $1.4_{- 0.4}^{+ 0.3}$ &  $7.05_{-0.73}^{+ 0.53}$    & $1.08 _{-0.16}^{+ 0.11}$ &  $0.30_{- 0.04}^{+ 0.06}$  &  $0.78\pm 0.11          $  &  $206_{- 23}^{+ 25}$&   $2.03 _{-0.12}^{+ 0.19}$    & 5.26  &  0.85  &  91       &  1.76   & 40/67   \\
36 & $1.6_{- 0.6}^{+ 0.4}$ &  $7.46_{-1.24}^{+ 0.73}$    & $1.07 _{-0.20}^{+ 0.15}$ &  $0.32_{- 0.05}^{+ 0.06}$  &  $0.88_{- 0.17}^{+ 0.12}$  &  $230_{- 39}^{+ 27}$&   $1.92 _{-0.14}^{+ 0.30}$    & 5.35  &  0.96  &  94       &  1.75   & 62/64   \\
37 & $3.3_{- 1.3}^{+ 1.3}$ &  $6.52_{-1.01}^{+ 1.11}$    & $1.04 _{-0.21}^{+ 0.22}$ &  $0.35_{- 0.07}^{+ 0.06}$  &  $0.98_{- 0.23}^{+ 0.26}$  &  $179_{- 42}^{+ 31}$&   $2.02 _{-0.23}^{+ 0.30}$    & 5.53  &  1.39  &  92       &  1.78   & 40/62   \\
38 & $3.2_{- 1.2}^{+ 0.6}$ &  $6.24_{-0.72}^{+ 0.67}$    & $0.96 _{-0.13}^{+ 0.14}$ &  $0.37\pm 0.05          $  &  $0.96_{- 0.18}^{+ 0.19}$  &  $199_\pm28        $&   $2.29 _{-0.18}^{+ 0.25}$    & 6.33  &  1.80  &  98       &  1.78   & 41/62   \\
\hline
\multicolumn{11}{l}{$^a$ Hydrogen column density, in units $10^{22}$ cm$^{-2}$.}\\
\multicolumn{11}{l}{$^b$ Normalization of the {\sf eqpair} model component.}\\
\multicolumn{12}{l}{$^c$ The unabsorbed total model flux, in units of $10^{-8} $erg cm$^{-2}$ s$^{-1}$.}\\
\multicolumn{12}{l}{$^d$ The unabsorbed model flux of the {\sf comptt} component, in units of $10^{-8}$ erg cm$^{-2}$ s$^{-1}$.}\\
\multicolumn{12}{l}{$^e$ Temperature of the emitting plasma in keV (for the {\sf eqpair} component).}\\
\multicolumn{12}{l}{$^f$ Photon spectral index of the {\sf eqpair} component in the 2--10 keV range.}\\
\end{tabular}
\end{center}
\label{tab:fits1999th}
\end{minipage}
\end{table*}

\subsubsection{High temperature of the optically-thick disc}

The additional component may, in principle, be emitted by the
hottest part of the optically-thick disc provided its temperature
is high enough. We find that the spectra of the observations 1--4,
18--38 can be well fitted with $\tin\sim 1$ keV. But \citet{ds01}
showed that the spectral decomposition of the \sax\/ data of Cyg
X-1 in the hard state requires the presence of both the soft
blackbody disc photons (with $\ktbb\simeq 0.1$--0.2 keV, and an
additional soft excess component. Therefore, we consider this
model to be not appropriate.

\subsubsection{Hybrid model}

We also tried to apply hybrid thermal+nonthermal {\sf eqpair}
model  to the spectra. Most of the data require nearly nonthermal
injection and the resulting spectrum becomes a power-law directed
by PCA part of the spectrum. Because the observed hard tail is not
power-law, large reflection appears to mimic the cutoff region,
but it is still not enough to describe both the soft excess below
10 keV and the hard tail, in which the systematic difference
between the data and the model remains. Therefore, we ruled out
this model.

\subsubsection{Two thermal Comptonization components}
\label{sec:twoth}

\begin{figure}
\centerline{\epsfig{file=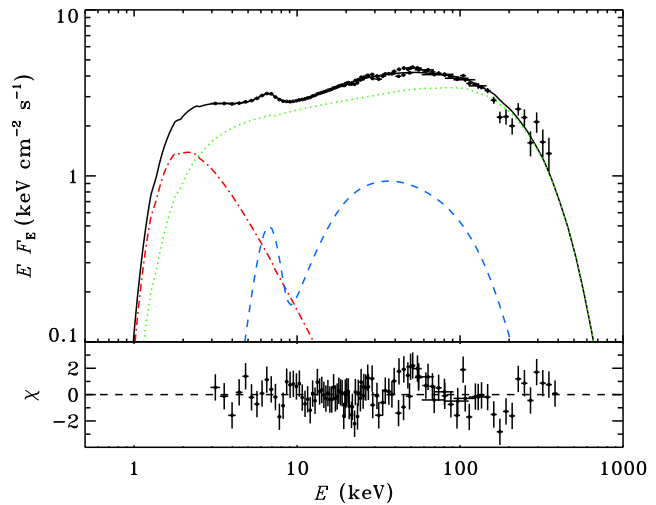,width=8.5cm}}
\caption{The model
spectrum with two thermal Comptonization components (model 1,
Sect.~\ref{sec:twoth}) fitted to the \xte+OSSE observation 24
(from 1999). The spectral components of the fit are shown by the
dotted (green), dot-dashed (red), and dashed (blue) curves, which
correspond to the main thermal-Comptonization continuum, the
additional thermal Comptonization with $kT_\mathrm{e}=20$ keV and
$\tau=1$, and the Compton reflection including the Fe line,
respectively. The solid (black) curve shows the total spectrum.
The lower panel shows the residuals of the fit.
}
\label{fig:thprimer}
\end{figure}

The soft component can be described by additional thermal
Comptonization \citep{fro01,ds01}. We use the model 1, {\sf
phabs(comptt+eqpair+gaussian)}, where {\sf eqpair} gives the main
Comptonization  and {\sf comptt} \citep{t94}  -- the additional
soft component. Since the parameters of the soft excess are rather
weakly constrained by our data, we fixed the parameters of {\sf
comptt} at $kT_\mathrm{e}=20$ keV and $\tau = 1$. Such a model
produces a soft power law that does not extend to very high
energies. This model fitted to the data of the observation 24 is
shown in Fig.~\ref{fig:thprimer}. In spite of its simplicity, it
yields a relatively good description of the data.

\begin{figure*}
\centerline{\epsfig{file=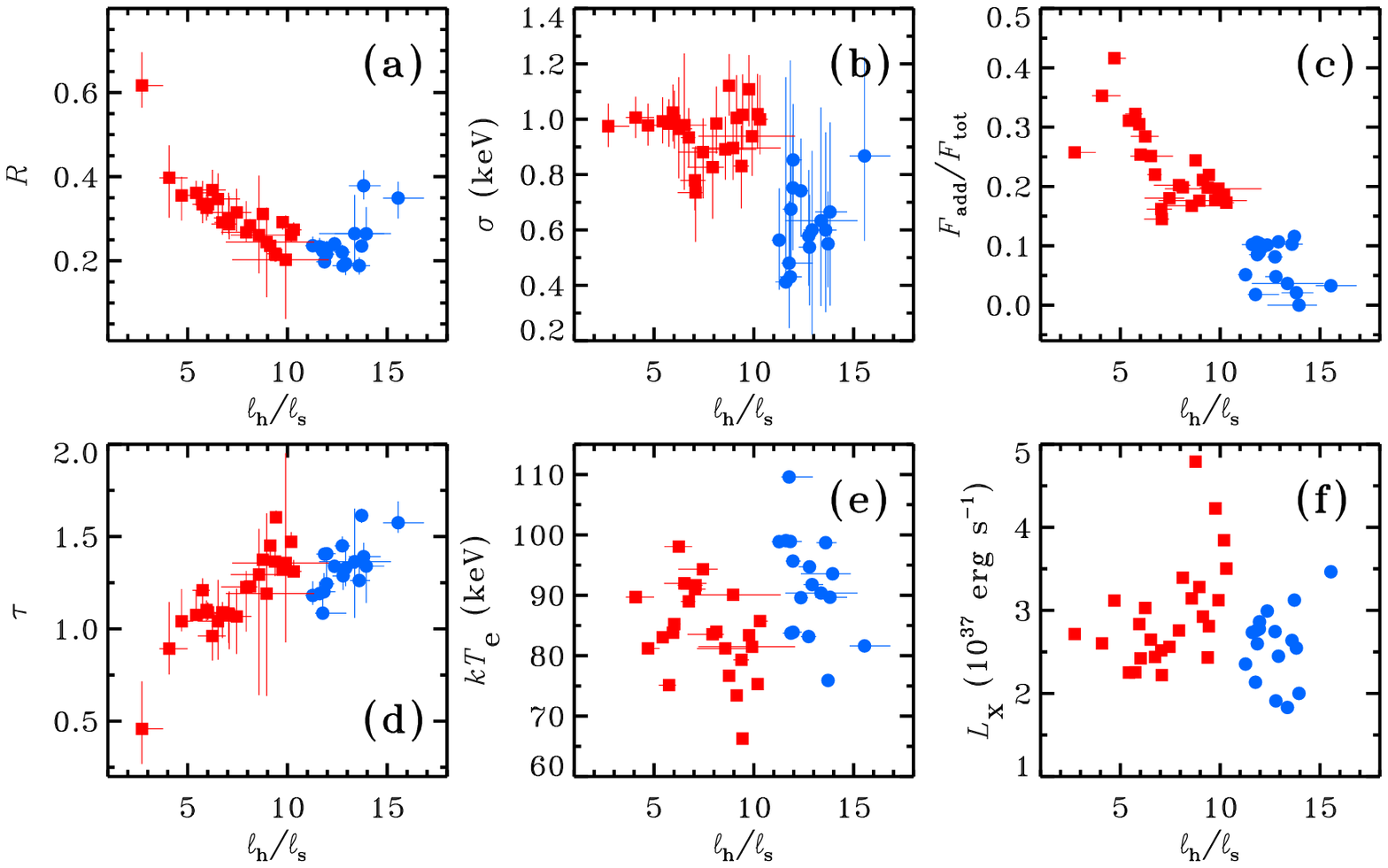,width=\textwidth}}
\caption{Dependencies of the spectral parameters for  model 1
(Sect.~\ref{sec:twoth})
fitted to the 1991 and 1997  (blue filled circles) and the 1996,
1998 and 1999 data (red filled squares) on
Compton amplification factor of the
main Comptonization component $\lhls$.
(a) The reflection fraction $\Refl$;
(b) the relativistic smearing Gaussian width $\sigma$ at
6.4 keV;  (c) ratio of the additional thermal
Comptonization flux to the total flux;
(d) Thomson optical depth of the main Comptonization continuum
component $\tau$; (e)  electron temperature of the
main Comptonization component $\kte$; (f) total
luminosity of the Comptonizing cloud (assuming $D=$2 kpc).
} \label{fig:thpic}
\end{figure*}

The fit parameters are given in Table \ref{tab:fits1999th}, and
the dependencies between various model parameters are shown in
Fig. \ref{fig:thpic}. We also quote the values of the
spectral index $\Gamma$ of the power law
obtained from the least-square fitting of the logarithm of the
intrinsic model flux at a logarithmic energy grid in the 2--10 keV
range (chosen to enable comparison with results of other papers).

The meaning of the normalization of {\sf eqpair} is described
in Sect.~\ref{sec:obs1999} above.
 Substituting $f_c=1$, $M=10\msun$, $i=50\degr$, $D=2.0$ kpc
(see references in G99, \citealt{fro01}) and $\betac=1.7$
\citep{st95}, we expect the normalization of $\simeq$1.92. The
lower normalization of the obtained fits (see
Table~\ref{tab:fits1999th}), is caused either by a larger
$\betac\sim 1.77$--1.98 or by a smaller covering factor $f_c$. A
slightly smaller, than the assumed disc temperature of 200 eV, can
also reduce the normalization. On the other hand, some flat state
normalizations are larger than the expected value of 1.92. This
cannot be explained by changing $f_c$ (smaller covering factor can
only reduce the normalization), but could be a result of somewhat
larger inner disc radius or larger temperature. The largest
observed normalizations correspond to $\ktbb \sim 220-245$ eV.

\begin{figure}
\centerline{\epsfig{file=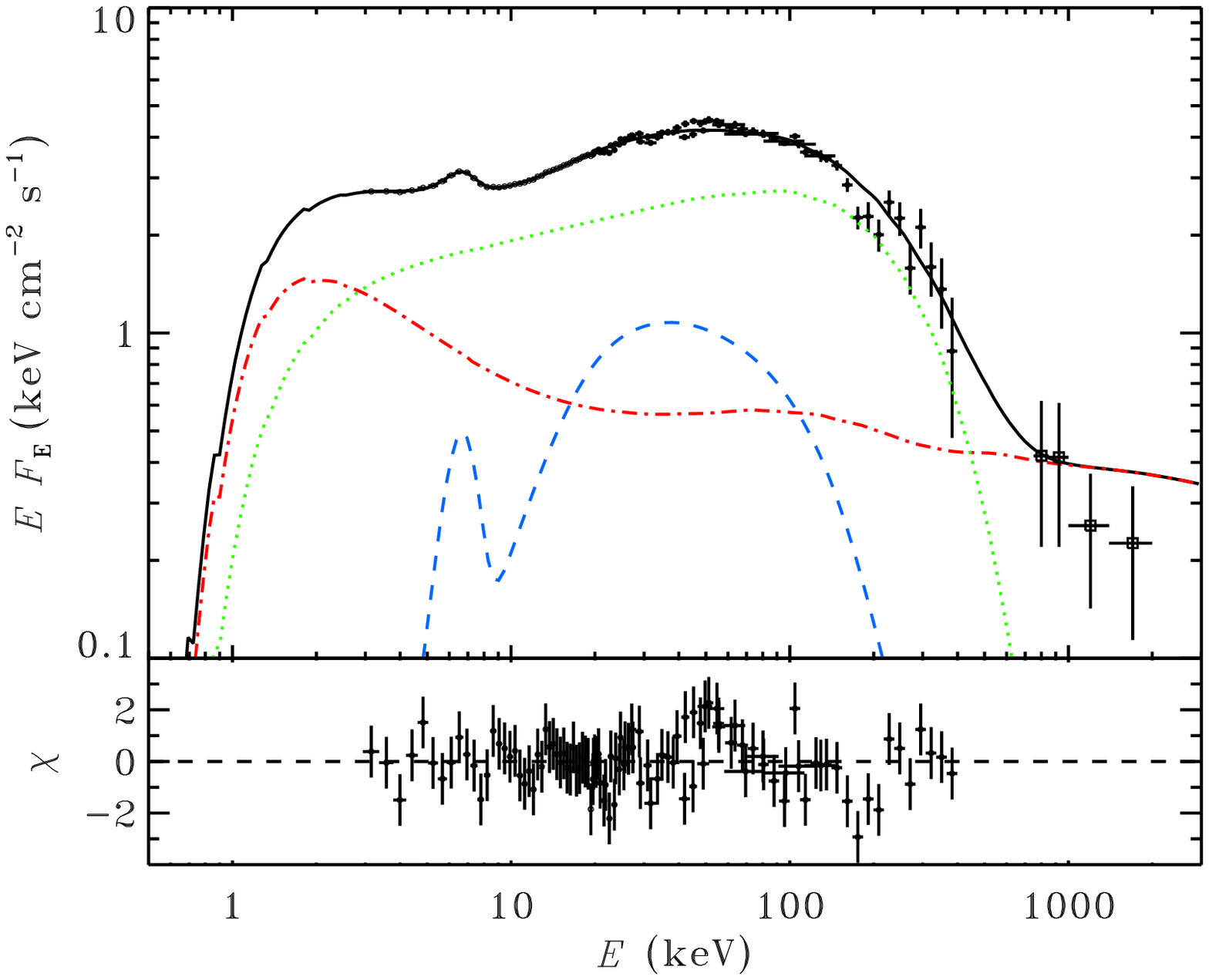,width=8.cm}}
\caption{ 
The model spectrum with the thermal and nonthermal Comptonization components (model 2,
Sect.~\ref{sec:thnth}) fitted to the \xte+OSSE observation 24
(from 1999) together with the COMPTEL
\citep{mccon02} hard-state data (marked by squares). The spectral components of
the fit are shown by the dotted (green), dot-dashed (red), and
dashed (blue) curves, which correspond to the main
thermal-Comptonization continuum, the additional nonthermal
Comptonization, and the Compton reflection including the Fe line,
respectively. The solid curve shows the total spectrum.
The lower panel shows the residuals of the fit.
}
\label{fig:nthprimer}
\end{figure}

We see that there is a clear anti-correlation between the strength of the soft
component and hardness of spectrum expressed in terms of $\lhls$.
There is also a correlation between $\tau$ and $\lhls$.
The reflection amplitude is correlated with the Fe line
equivalent width $EW$ and anti-correlated with $\lhls$
(while at high values of $\lhls$ the anti-correlation
possibly breaks down).

The electron temperature $\kte$ is a calculated parameter and no
errors on it can be obtained from fitting. However, we estimated
its $1\sigma$ limits using its extremal values within the
uncertainties of the parameters controlling spectral shape, i.e.,
$\lhls$ and $\tau$. This estimation gives us a possible conservative
error on $\kte$ of about 15 keV for both models 1 and 2. Taking this
into account, no correlations between $\kte$--flux and
$\kte-\lhls$ are apparent.

\begin{figure*}
\centerline{\epsfig{file=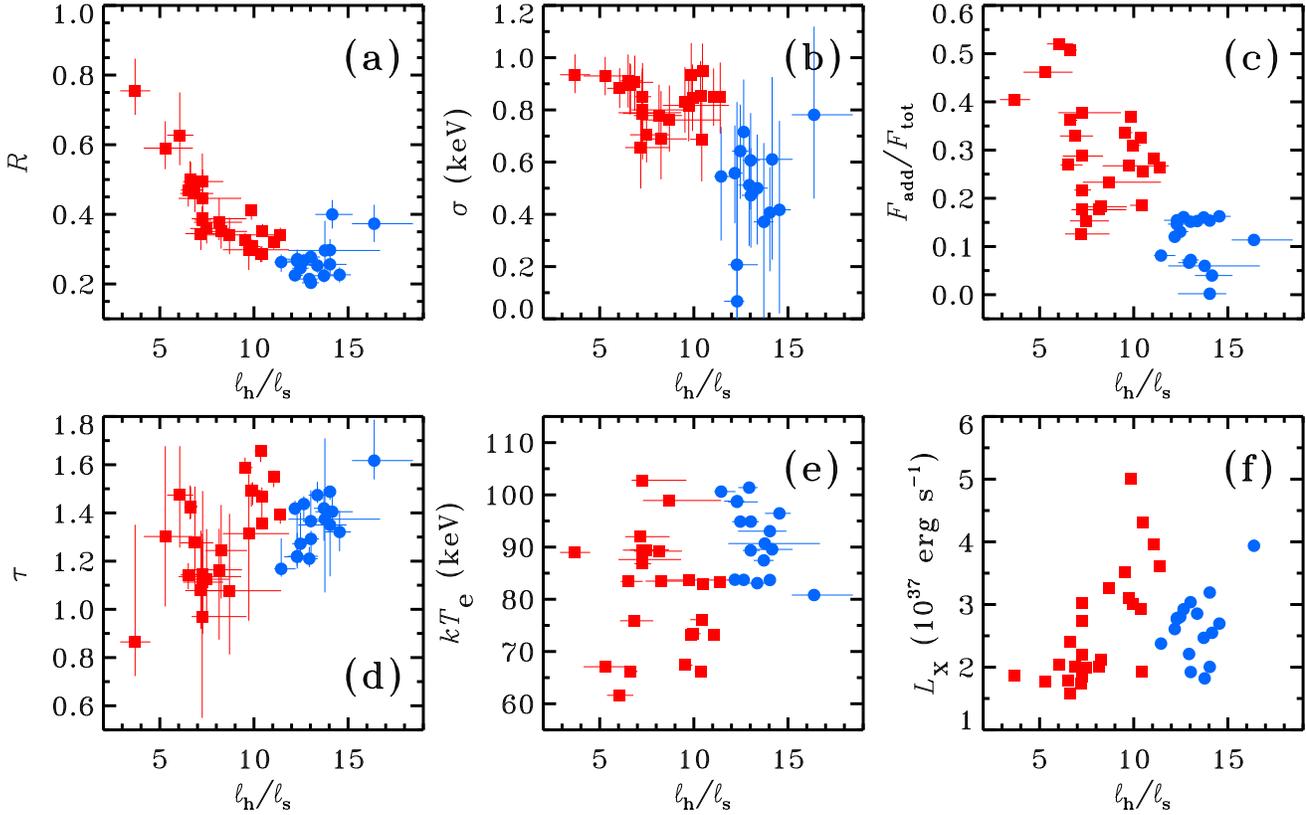,width=\textwidth}}
\caption{Dependencies of the spectral parameters for  model 2
(Sect.~\ref{sec:thnth})
fitted to the 1991 and 1997  (blue filled circles) and the 1996,
1998 and 1999 data (red filled squares) on
Compton amplification factor of the
main Comptonization component $\lhls$.
The meaning of the axes is the same as in Fig.~\ref{fig:thpic},
except (c) ratio of the additional nonthermal
Comptonization flux in model 2 to the total flux. }
\label{fig:nthpic}
\end{figure*}

\subsubsection{Thermal continuum and nonthermal Comptonization component}
\label{sec:thnth}

As an alternative to the second thermal-Compton component, we
consider here addition of a {\em nonthermal} Comptonization
component. We find that such a nonthermal component can describe
both the soft excess and the MeV nonthermal tail observed in hard
states by COMPTEL \citep{mccon02}, while neither of these
components can be described by the main thermal-Compton emission.
This model fitted to the spectrum 24 is shown in
Fig.~\ref{fig:nthprimer}. Note that the COMPTEL data are shown for
illustration only and were not taken into account in the fitting.
The used model 2 consists of {\sf phabs(eqpair+eqpair+gaussian)},
in which the second {\sf eqpair} component produces the nonthermal
spectrum. In {\sf eqpair}, the available power is supplied in part
into heating electrons and in part into their acceleration, with
the resulting steady-state electron distribution calculated
self-consistently. The compactness corresponding to the
acceleration is hereafter denoted as $\lnth$. Then the relative
fraction of the input power going into the nonthermal acceleration
is $\lnth/\lh$, where $\lh$ (as before) corresponds to the total
rate of energy dissipation in the plasma.

For that additional component, we assumed that all the available power goes into nonthermal
acceleration, i.e., $\lnth/\lh=1$. Note that the resulting self-consistent
electron distribution is
not purely nonthermal but hybrid, i.e., it does contain a low-energy Maxwellian
heated by Compton and Coulomb interactions. We further assume $\Refl=0$, the
power-law index of the accelerated electrons of $\ginj=2.4$ (see \citealt{mccon02}; G99;
\citealt{fro01,pc98}), the minimum and maximum
Lorentz factors of the power law of $\gamma_\mathrm{min}=1.3$ and
$\gamma_\mathrm{max}=1000$, respectively, $\ktbb$ equal to that of the main
component, $\lhls=1$, and $\taub=1$. No pair production is required, and $\taut$
is found to be equal to $\taub$.

\begin{figure}
\centerline{\epsfig{file=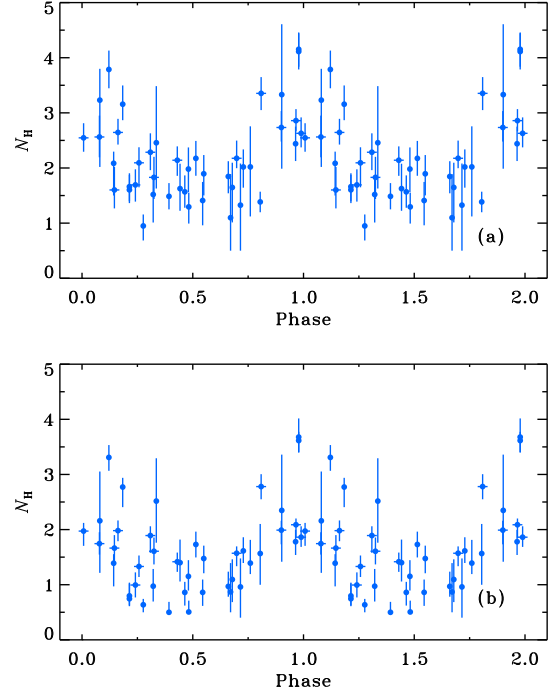,width=7.7cm}}
\caption{ (a)
Fitted value of absorption column density $\nh$ for model 1 vs.
orbital phase of the binary system. (b) Same, for model 2.
Distributions are repeated twice for clarity. $\nh$ is in units of
$10^{22}$ cm$^{-2}$.} \label{fig:nh}
\end{figure}

The best-fit parameters are presented in Table~\ref{fits1999nth}
and correlations between them are shown in Figs.\ \ref{fig:nthpic}
and \ref{fig:old_vs_new}. The normalization of the thermal
{\sf eqpair} component corresponds to $\betac=1.63$--$2.01$,
the values similar to those obtained with model 1.
The strength of the additional (non-thermal) component is again
clearly anti-correlated with the spectral hardness.
The anti-correlation between $\Refl$ and $\lhls$ is apparent and corresponds to the
$\Refl-\Gamma$ correlation that we discuss in Sect.~\ref{sec:old_vs_new}.
The values of $\Refl$ and the Fe line
equivalent width $EW$ are well correlated and can be approximated
by a linear function passing through zero (see Fig.~\ref{fig:old_vs_new}c).
Again, there is a correlation between $\taut$ and $\lhls$.
The electron temperature does not seemingly depend on the hardness, but
the spread becomes smaller at larger $\lhls$ where
$\kte\sim 90$ keV.

\begin{table*}
\begin{minipage}{180mm}\caption{The best-fit parameters for model 2.}
\fontsize{8pt}{4mm} \selectfont
\begin{center}
\begin{tabular}{ccccccccccccc}
\hline Obs. &$\nh^a$& $\lhls$                &  $\taut$        & $\Refl$         &$\sigma$ & $EW$ &$N^b$ &$F_\mathrm{tot}^c$ & $F_\mathrm{add}^d$  &$\kte^e$ &  $\Gamma^f$ & $\chi^2/{\rm dof}$ \\
\hline
G1 &   $ 0.9 _{-0.4}^{+ 0.5} $ & $  14.2_{- 0.9}^{+ 1.1} $ & $  1.41\pm  0.08           $ & $ 0.40 \pm0.04           $ & $0.61_{- 0.38}^{+ 0.32}   $ & $  182_{- 52}^{+ 59} $& $1.29\pm 0.09            $  &   5.32 & 0.21 & 90 & 1.62 &  49/73     \\
G2 &   $ 1.0 _{-0.6}^{+ 0.5} $ & $  16.4_{- 1.2}^{+ 2.1} $ & $  1.62_{-  0.08}^{+ 0.17} $ & $ 0.37 \pm0.05           $ & $0.78_{- 0.32}^{+ 0.34}   $ & $  210_{- 60}^{+ 68} $& $1.50_{- 0.20}^{+ 0.12}  $  &   8.23 & 0.93 & 81 & 1.58 &  56/74     \\
G3 &   $ 1.4 _{-0.2}^{+ 0.4} $ & $  14.0_{- 1.7}^{+ 0.9} $ & $  1.35_{-  0.21}^{+ 0.08} $ & $ 0.26_{- 0.03}^{+ 0.07} $ & $0.58_{- 0.58}^{+ 0.93}   $ & $  144_{- 50}^{+ 57} $& $1.12_{- 0.06}^{+ 0.10}   $ &   4.19 & 0.01 & 93 & 1.62 &  53/73     \\
G4 &   $ 1.6 _{-0.6}^{+ 0.5} $ & $  13.8_{- 1.9}^{+ 2.9} $ & $  1.37_{-  0.30}^{+ 0.34} $ & $ 0.30_{- 0.08 }^{+0.09} $ & $0.56_{- 0.58}^{+ 0.95}   $ & $  122_{- 48}^{+ 55} $& $0.97_{- 0.15}^{+ 0.13}   $ &   3.81 & 0.23 & 91 & 1.62 &  53/73     \\
01 &   $ 2.0 \pm0.2          $ & $  11.4\pm 0.3           $ & $ 1.39_{-  0.04}^{+ 0.02} $ & $ 0.34\pm 0.02           $ & $0.85_{- 0.14}^{+ 0.13}   $ & $  148_{- 21}^{+ 19} $& $1.63_{- 0.05}^{+ 0.07}  $  &   7.55 & 1.98 & 83 & 1.66 &  498/478    \\
02 &   $ 1.4 \pm0.2          $ & $  10.5\pm 0.2           $ & $ 1.36\pm  0.02           $ & $ 0.35_{- 0.01}^{+ 0.02} $ & $0.95\pm 0.10             $ & $  203_{- 19}^{+ 21} $& $2.12\pm 0.06            $  &   9.00 & 2.30 & 83 & 1.67 &  528/493    \\
03 &   $ 1.6 _{-0.2}^{+ 0.1} $ & $  11.1_{- 0.1}^{+ 0.2} $ & $  1.55_{-  0.04}^{+ 0.03} $ & $ 0.32\pm 0.02           $ & $0.85\pm 0.11             $ & $  182_{- 17}^{+ 22} $& $1.81_{- 0.04}^{+ 0.05}  $  &   8.28 & 2.34 & 73 & 1.66 &  527/506    \\
04 &   $ 2.1 _{-0.2}^{+ 0.1} $ & $  9.86_{-0.24}^{+0.19}  $ &$  1.49_{-  0.07}^{+ 0.03} $ & $ 0.41_{- 0.03}^{+ 0.02} $ & $0.93\pm 0.12             $ & $  167_{- 17}^{+ 23} $& $2.16_{- 0.06}^{+ 0.09}  $  &  10.45 & 3.86 & 73 & 1.69 &  515/493    \\
05 &   $ 1.7 _{-0.3}^{+ 0.2} $ & $  13.0_{- 0.6}^{+ 0.5} $ & $  1.21_{-  0.04}^{+ 0.05} $ & $ 0.21\pm 0.02           $ & $0.51_{- 0.23}^{+ 0.21}   $ & $  81 _{- 17}^{+ 18} $& $1.17_{- 0.02}^{+ 0.05}  $  &   4.62 & 0.31 & 101&  1.63&  60/73     \\
06 &   $ 1.6 \pm0.2          $ & $  13.0_{- 0.5}^{+ 0.4} $ & $  1.29_{-  0.07}^{+ 0.03} $ & $ 0.20\pm 0.01           $ & $0.47_{- 0.20}^{+ 0.18}   $ & $  90 _{- 16}^{+ 18} $& $1.01_{- 0.03}^{+ 0.04}  $  &   4.02 & 0.29 & 95 & 1.63 &  372/402    \\
07 &   $ 1.5 _{-0.3}^{+ 0.2} $ & $  12.5\pm 0.4          $ & $  1.27_{-  0.08}^{+ 0.13} $ & $ 0.25\pm 0.02           $ & $0.64\pm 0.18             $ & $  108 \pm 21       $ & $1.42_{- 0.08}^{+ 0.06}  $ &   5.85 & 0.76 & 95 & 1.64 &  366/415    \\
08 &   $ 1.6 \pm0.2          $ & $  12.2_{- 0.3}^{+ 0.4} $ & $  1.42\pm  0.03           $ & $ 0.23\pm 0.02           $ & $0.56_{- 0.19}^{+ 0.18}   $ & $  106_{- 19}^{+ 20} $& $1.37_{- 0.05}^{+ 0.04}  $  &   5.45 & 0.65 & 84 & 1.64 &  401/415    \\
09 &   $ 1.9 _{-0.3}^{+ 0.2} $ & $  12.7_{- 0.2}^{+ 0.3} $ & $  1.44_{-  0.07}^{+ 0.03} $ & $ 0.27\pm 0.02           $ & $0.72_{- 0.19}^{+ 0.20}   $ & $  110_{- 18}^{+ 27} $& $1.40_{- 0.04}^{+ 0.06}  $  &   6.11 & 0.98 & 84 & 1.64 &  348/415    \\
10 &   $ 1.4 _{-0.3}^{+ 0.4} $ & $  11.5_{- 0.4}^{+ 0.7} $ & $  1.17_{-  0.03}^{+ 0.13} $ & $ 0.26_{- 0.03}^{+ 0.02} $ & $0.55_{- 0.25}^{+ 0.19}   $ & $  103_{- 26}^{+ 21} $& $1.37_{- 0.08}^{+ 0.05}  $  &   4.97 & 0.40 & 101&  1.66&  301/334    \\
11 &   $ 3.7 _{-0.3}^{+ 0.1} $ & $  12.3_{- 0.7}^{+ 0.4} $ & $  1.22_{-  0.05}^{+ 0.10} $ & $ 0.27_{- 0.02}^{+ 0.03} $ & $0.07_{- 0.07}^{+ 0.56}   $ & $  43_{- 10}^{+ 19} $ & $1.38_{- 0.06}^{+ 0.07}   $&   5.82 & 0.90 & 99 & 1.64 &  193/236    \\
12 &   $ 3.6 _{-0.2}^{+ 0.4} $ & $  12.3_{- 0.4}^{+ 1.1 }$ & $  1.22_{-  0.04}^{+ 0.18} $ & $ 0.27\pm 0.03           $ & $0.21_{- 0.21}^{+ 0.62}   $ & $  46_{- 12}^{+ 21} $ & $1.38_{- 0.06}^{+ 0.04}   $&   5.78 & 0.85 & 99 & 1.64 &  201/238    \\
13 &   $ 2.5\pm 0.8          $ & $  13.4_{- 0.5}^{+ 0.6 }$ & $  1.47_{-  0.03}^{+ 0.06} $ & $ 0.25\pm 0.02           $ & $0.50_{- 0.22}^{+ 0.21}   $ & $  88 _{- 22}^{+ 21} $& $1.31_{- 0.07}^{+ 0.06}  $  &   5.96 & 0.91 & 83 & 1.62 &  365/411    \\
14 &   $ 1.7\pm 0.2          $ & $  13.0_{- 0.4}^{+ 0.5 }$ & $  1.37_{-  0.03}^{+ 0.13} $ & $ 0.28\pm 0.02           $ & $0.61_{- 0.19}^{+ 0.18}   $ & $  109_{- 19}^{+ 20} $& $1.42_{- 0.06}^{+ 0.05}  $  &   6.35 & 0.96 & 89 & 1.63 &  391/415    \\
15 &   $ 2.8 _{-0.4}^{+ 0.2} $ & $  14.6_{- 0.8}^{+ 0.6 }$ & $  1.32_{-  0.08}^{+ 0.09} $ & $ 0.23\pm 0.02           $ & $0.42_{- 0.40}^{+ 0.34}   $ & $  62 _{- 17}^{+ 27} $& $1.14_{- 0.04}^{+ 0.07}  $  &   5.63 & 0.92 & 96 & 1.61 &  198/244    \\
16 &   $ 2.8\pm 0.2          $ & $  14.0\pm 0.3          $ & $  1.49_{-  0.03}^{+ 0.02} $ & $ 0.30_{- 0.02}^{+ 0.01} $ & $0.41_{- 0.22}^{+ 0.17}   $ & $  85 _{- 15}^{+ 16} $& $1.38\pm 0.04            $  &   6.67 & 1.02 & 84 & 1.62 &  436/402    \\
17 &   $ 3.3\pm 0.2          $ & $  13.7\pm 0.5          $ & $  1.42_{-  0.13}^{+ 0.08} $ & $ 0.22\pm 0.02           $ & $0.37_{- 0.37}^{+ 0.30}   $ & $  58 _{- 15}^{+ 19} $& $1.11_{- 0.04}^{+ 0.05}  $  &   5.16 & 0.83 & 87 & 1.62 &  342/415    \\
18 &   $ 2.0 _{-0.3}^{+ 0.1} $ & $  9.53_{-0.23}^{+0.40}  $ &$  1.59_{-  0.03}^{+ 0.04} $ & $ 0.33\pm 0.02           $ & $0.83_{- 0.12}^{+ 0.13}   $ & $  150_{- 16}^{+ 32} $& $1.70_{- 0.06}^{+ 0.07}  $  &   7.34 & 2.47 & 67 & 1.69 &  481/489    \\
19 &   $ 2.0 _{-0.3}^{+ 0.2} $ & $  10.4_{- 0.3}^{+ 0.1 }$ & $  1.66_{-  0.05}^{+ 0.01} $ & $ 0.29\pm 0.02           $ & $0.86_{- 0.15}^{+ 0.13}   $ & $  148_{- 21}^{+ 22} $& $1.34_{- 0.03}^{+ 0.06}  $  &   6.10 & 1.99 & 66 & 1.67 &  455/502    \\
20 &   $ 1.9 \pm0.2          $ & $  9.93_{-0.18}^{+ 0.43} $ &$  1.49_{-  0.06}^{+ 0.03} $ & $ 0.31_{- 0.02}^{+ 0.01} $ & $0.85\pm 0.13             $ & $  151_{- 20}^{+ 19} $& $1.47_{- 0.05}^{+ 0.04}  $  &   6.30 & 1.95 & 73 & 1.69 &  467/502    \\
21 &   $ 1.7 \pm0.2          $ & $  9.73_{-1.38}^{+ 2.14} $ &$  1.31_{-  0.36}^{+ 0.25} $ & $ 0.30_{- 0.06}^{+ 0.04} $ & $0.82\pm 0.14             $ & $  139\pm 20 $        & $1.64_{- 0.46}^{+ 0.30}  $  &   6.47 & 1.73 & 84 & 1.69 &  47/73      \\
22 &   $ 1.3\pm 0.2          $ & $  8.70_{-1.41}^{+ 2.75} $ &$  1.08_{-  0.26}^{+ 0.32} $ & $ 0.34_{- 0.06}^{+ 0.05} $ & $0.76_{- 0.12}^{+ 0.13}   $ & $  160_{- 20}^{+ 30} $& $1.98_{- 0.47}^{+ 0.39}  $  &   6.82 & 1.59 & 99 & 1.71 &  44/73      \\
23 &   $ 1.2\pm 0.3          $ & $  7.25_{-0.56}^{+ 2.36} $ &$  0.97_{-  0.42}^{+ 0.31} $ & $ 0.39\pm 0.04           $ & $0.80\pm 0.12             $ & $  171_{- 21}^{+ 22} $& $2.29_{- 0.46}^{+ 0.34}  $  &   6.32 & 1.11 & 103&  1.75&  43/73      \\
24 &   $ 0.7 _{-0.1}^{+ 0.3} $ & $  6.61_{-0.31}^{+ 0.37} $ &$  1.42_{-  0.05}^{+ 0.09} $ & $ 0.49_{- 0.04}^{+ 0.06} $ & $0.89_{- 0.11}^{+ 0.08}   $ & $  301_{- 39}^{+ 22} $& $1.15_{- 0.09}^{+ 0.06}  $  &   5.03 & 2.55 & 66 &  1.78&  379/409    \\
25 &   $ 0.5 _{-0  }^{+0.2}  $ & $  6.06_{-0.67}^{+ 0.73} $& $  1.47_{-  0.22}^{+ 0.20} $ & $ 0.63_{- 0.09}^{+ 0.12} $ & $0.88_{- 0.07}^{+ 0.08}   $ & $  346_{- 27}^{+ 26} $& $1.17_{- 0.14}^{+ 0.07}  $  &   4.28 & 2.22 & 62 &  1.80&  354/409    \\
26 &   $ 0.5 _{-0  }^{+ 0.2} $ & $  5.29_{-1.15}^{+ 1.47} $ &$  1.30_{-  0.29}^{+ 0.38} $ & $ 0.59_{- 0.06}^{+ 0.08} $ & $0.93\pm 0.07             $ & $  348 \pm 27  $      & $1.32_{- 0.28}^{+ 0.35}  $    &   3.69 & 1.70 & 67 &  1.83&  52/67      \\
27 &   $ 1.0 _{-0.2}^{+ 0.3} $ & $  6.86_{-0.78}^{+ 0.98} $ &$  1.28_{-  0.17}^{+ 0.23} $ & $ 0.46_{- 0.05}^{+ 0.06} $ & $0.91_{- 0.11}^{+ 0.10}   $ & $  263_{- 29}^{+ 28} $& $1.48_{- 0.14}^{+ 0.21}  $  &   4.21 & 1.39 & 76 &  1.77&  380/409    \\
28 &   $ 0.8\pm 0.2          $ & $  6.63_{-0.31}^{+ 0.36} $ &$  1.43_{-  0.05}^{+ 0.08} $ & $ 0.50_{- 0.04}^{+ 0.05} $ & $0.90_{- 0.11}^{+ 0.08}   $ & $  299_{- 35}^{+ 23} $& $1.14_{- 0.09}^{+ 0.07}  $  &   3.31 & 1.20 & 66 &  1.78&  379/409    \\
29 &   $ 0.9\pm 0.2          $ & $  6.51_{-0.39}^{+ 0.77} $ &$  1.14_{-  0.06}^{+ 0.05} $ & $ 0.47_{- 0.05}^{+ 0.04} $ & $0.91\pm 0.10             $ & $  264\pm 27 $        & $1.51_{- 0.17}^{+ 0.13}  $  &   3.74 & 1.01 & 83 & 1.78 &  388/409    \\
30 &   $ 0.6\pm 0.1          $ & $  3.68_{-0.79}^{+ 0.81} $ &$  0.86_{-  0.14}^{+ 0.49} $ & $ 0.75_{- 0.07}^{+ 0.09} $ & $0.93_{- 0.07}^{+ 0.08}   $ & $  324_{- 23}^{+ 26} $& $2.17_{- 0.73}^{+ 0.65}  $  &   3.91 & 1.58 & 89 & 1.91 &  49/67      \\
31 &   $ 1.0\pm 0.2          $ & $  7.26_{-0.41}^{+ 0.47} $ &$  1.15_{-  0.12}^{+ 0.14} $ & $ 0.38\pm 0.04           $ & $0.85\pm 0.11             $ & $  228_{- 25}^{+ 26} $& $1.53_{- 0.15}^{+ 0.11}  $  &   3.86 & 0.84 & 87 & 1.75 &  388/409    \\
32 &   $ 1.8\pm 0.2          $ & $  10.4_{-0.6 }^{+  0.3 }$ & $ 1.47_{-  0.13}^{+ 0.04} $ & $ 0.29\pm 0.02           $ & $0.69_{- 0.16}^{+ 0.15}   $ & $  128_{- 21}^{+ 22} $& $1.19_{- 0.04}^{+ 0.11}  $  &   4.03 & 0.75 & 76 & 1.67 &  388/409    \\
33 &   $ 1.4 _{-0.4}^{+ 0.2} $ & $  8.26_{-0.82}^{+ 1.38} $ &$  1.24_{-  0.20}^{+ 0.19} $ & $ 0.35_{- 0.05}^{+ 0.04} $ & $0.69_{- 0.15}^{+ 0.18}   $ & $  151_{- 21}^{+ 38} $& $1.63\pm 0.21            $  &   4.44 & 0.81 & 83 & 1.72 &  62/67      \\
34 &   $ 1.0\pm 0.3          $ & $  7.18_{-0.83}^{+ 1.54} $ &$  1.08_{-  0.16}^{+ 0.25} $ & $ 0.34_{- 0.05}^{+ 0.03} $ & $0.66_{- 0.16}^{+ 0.17}   $ & $  160_{- 26}^{+ 30} $& $1.66_{- 0.30}^{+ 0.10}  $  &   3.65 & 0.46 & 92 & 1.75 &  42/67      \\
35 &   $ 0.9\pm 0.2          $ & $  7.48_{-0.87}^{+ 1.20} $ &$  1.13_{-  0.15}^{+ 0.21} $ & $ 0.36_{- 0.04}^{+ 0.05} $ & $0.70\pm 0.11             $ & $  189_{- 22}^{+ 23} $& $1.74_{- 0.24}^{+ 0.23}  $  &   4.15 & 0.63 & 89 & 1.75 &  41/67      \\
36 &   $ 1.1\pm 0.4          $ & $  8.15_{-1.33}^{+ 1.22} $ &$  1.16_{-  0.29}^{+ 0.17} $ & $ 0.38_{- 0.05}^{+ 0.07} $ & $0.78_{- 0.11}^{+ 0.12}   $ & $  203_{- 25}^{+ 26} $& $1.56_{- 0.21}^{+ 0.31}  $  &   4.21 & 0.74 & 89 & 1.73 &  58/64      \\
37 &   $ 2.3 _{-0.9}^{+ 1.0} $ & $  7.27_{-1.28}^{+ 2.05} $ &$  1.14_{-  0.24}^{+ 0.35} $ & $ 0.45_{- 0.06}^{+ 0.08} $ & $0.79_{- 0.18}^{+ 0.19}   $ & $  149_{- 27}^{+ 29} $& $1.54_{- 0.18}^{+ 0.35}  $  &   5.72 & 2.16 & 88 & 1.75 &  41/62      \\
38 &   $ 2.2\pm 0.9          $ & $  7.26_{-1.02}^{+ 1.12} $ &$  1.11_{-  0.19}^{+ 0.20} $ & $ 0.49_{- 0.06}^{+ 0.08} $ & $0.79_{- 0.13}^{+ 0.16}   $ & $  169_{- 24}^{+ 25} $& $1.60_{- 0.25}^{+ 0.34}  $  &   4.58 & 1.32 & 89 & 1.75 &  45/62      \\
\hline
\multicolumn{11}{l}{$^a$ Hydrogen column density, in units $10^{22}$ cm$^{-2}$.}\\
\multicolumn{12}{l}{$^b$ Normalization of the thermal {\sf eqpair} model component.}\\
\multicolumn{12}{l}{$^c$ The unabsorbed total model flux (without reflection) in units of $10^{-8}$
erg cm$^{-2}$ s$^{-1}$.}\\
\multicolumn{12}{l}{$^d$ The unabsorbed model flux from the nonthermal component, in units of $10^{-8}$   erg cm$^{-2}$ s$^{-1}$.}\\
\multicolumn{12}{l}{$^e$ Temperature of the emitting plasma in keV (for the thermal {\sf eqpair} component).}\\
\multicolumn{12}{l}{$^f$ Photon spectral index of the thermal {\sf eqpair} component in the 2--10 keV range.}\\
\end{tabular}
\end{center}
 \label{fits1999nth}
\end{minipage}
\end{table*}

\section{Discussion}

\subsection{Absorption}

As noted in Sect.~\ref{sec:analysis}, the
hydrogen column density $\nh$ was free in our fits.
We find that our data require absorption significantly
larger than $0.6\pm0.2\times 10^{22}$ cm$^{-2}$ which is derived from the
reddening towards the companion star \citep{bc95}.

Cyg X-1 is known to show X-ray dips in its
light curve, caused by obscuration by the stellar wind from
the companion star. During dips the absorption increases
up to $\nh=20\times 10^{22}$  cm$^{-2}$.
The absorption  shows strong orbital phase dependence with maximal
column density around phase 0, when companion star is in front of the
black hole \citep{feng02,bc00}.
We plotted the fitted values of $\nh$
versus Cyg X-1 orbital phase (see Fig.~\ref{fig:nh}),
using the ephemeris with the time of the primary minimum at
$50234.79$ MJD and the period of
$5.599829$ days \citep{broc99,lasala98}.
We see a good correlation of $\nh$ with the orbital phase, which
indicates that variable absorption indeed can be
caused by the companion's wind obscuring the X-rays from the black hole.

We note that since our
models are relatively complicated in the range of 3--10 keV
(consisting of absorption, Comptonization together with the seed
photon emission and a soft excess), and also, the PCA energy range
is affected by absorption rather weakly, it is
difficult to determine the exact values of $\nh$. Even
though,  the derived values of the parameter are in the range quoted by
other authors and its relative changes are quite remarkable.

\subsection{Spectral variability patterns}

Using \gro/BATSE and \xte/ASM data, \citet{zdz02} showed that Cyg
X-1 has two types of variability -- changes of flux without spectral
slope change and pivoting at $\sim$ 50 keV which produces
anti-correlation of the fluxes in the soft and hard part of the
spectrum. However,  those instruments do not provide detailed
spectral information giving    fluxes only in some energy
intervals. The effective photon spectral indices using fluxes in
two energy bands and calculated in the wide (20-300 keV) energy
interval, where real spectra experience a cutoff, should be
treated with caution. Now we have a possibility to check the
results of \citet{zdz02}  using our set of observations, on which
we have a detailed spectral information in the wide energy range.
For this purpose we are using the model spectra obtained from
model 1.

\begin{figure}
\centerline{\epsfig{file=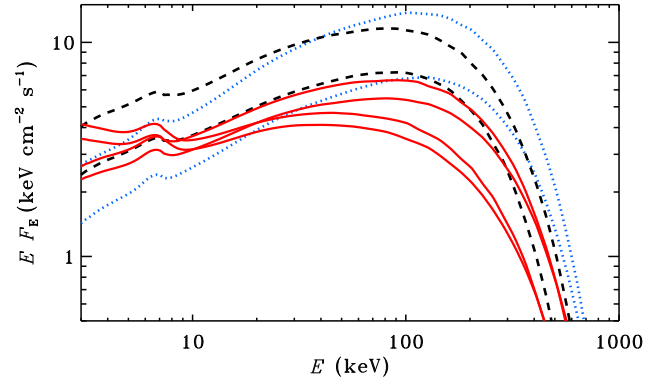,width=8.5cm}} \caption{Sample of
the spectra from different periods. Lowest and highest spectra
from each time period are shown. Red solid lines -- 1999; blue
dotted lines -- 1991 (the 1997 spectra are similar to them); black
dashed lines -- 1996 and 1998. The 1991+1997 and 1996+1998 data
form two groups with slightly different spectral slopes. }
\label{fig:pivotstat}
\end{figure}

In Fig.~\ref{fig:pivotstat} we present the spectra related to
different time periods. It is possible to see that  1991, 1996 --
1998 spectra only change their normalization. However, spectral
slope is different for various years and forms  two groups:
1991+1997 spectra and 1996+1998 ones. In the 1999 data, we again
can see normalization changes (see pairs of red solid curves) as
well as pivoting behaviour.

\begin{figure*}
\centerline{\epsfig{file=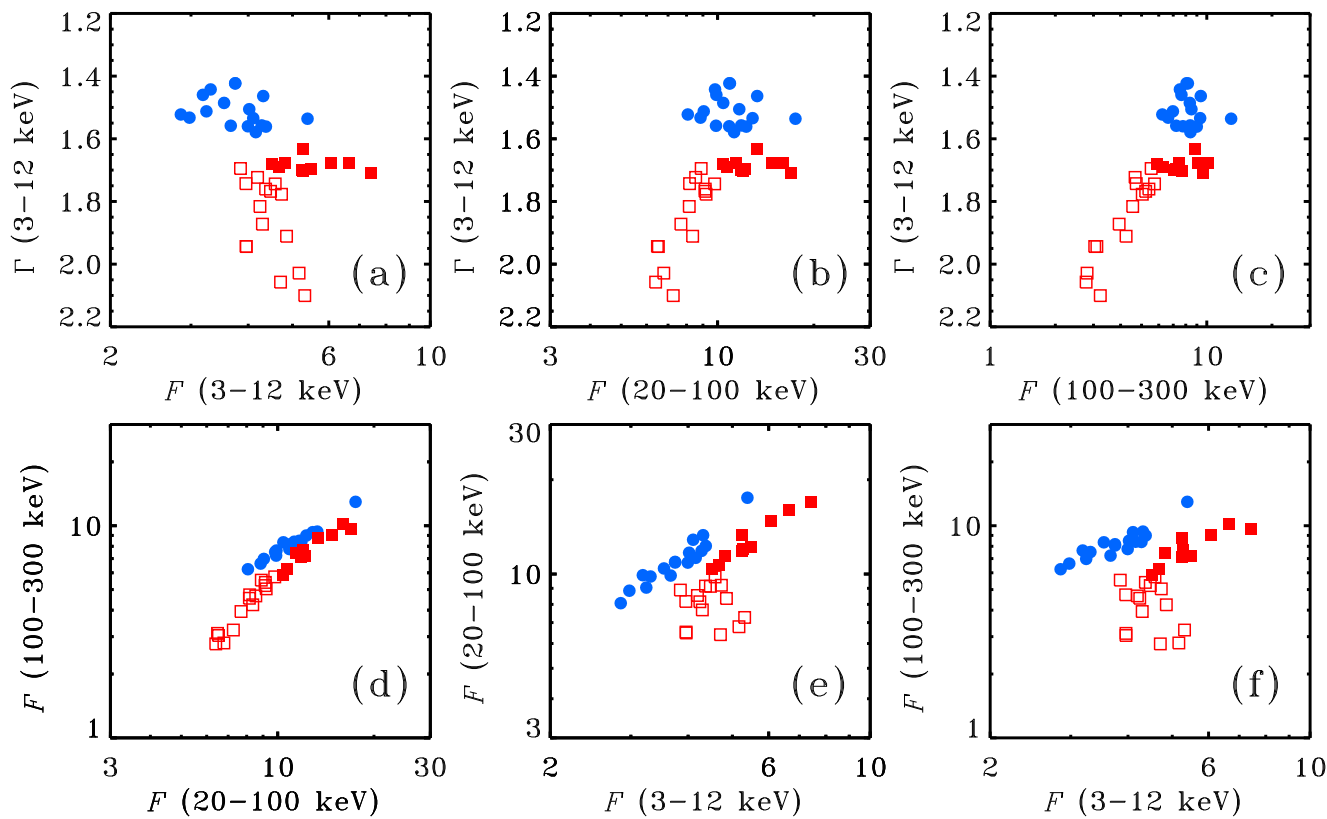,width=\textwidth}}
\caption{Flux--spectral index (3--12 keV) and flux--flux
correlations. Respective energy intervals are indicated along the
axis. The 1991 and 1997 data indicated by blue filled circles,
1996 and 1998 -- by red  filled squares and 1999 -- by red open
squares.
All fluxes are in units of keV cm$^{-2}$ s$^{-1}$. }
\label{fig:pivoting}
\end{figure*}

The $\Gamma$--flux correlations are shown on
Fig.~\ref{fig:pivoting}. It is clearly seen that the 1991 and 1996--1998
data do not show   dependence between flux and $\Gamma$. The 1999
data show clear anti-correlation between $\Gamma$  and flux on low
energies (3--12 keV) and correlation on high energies (20-100,
100-300 keV), that indicates the pivoting behaviour with the pivot
energy between 12 and $\sim$ 50 keV. On
Fig.~\ref{fig:pivoting}d (see also fig. 8 in \citealt{zdz02}), it
can be seen that spectra from 1996 and 1998 are somewhat softer
at high energies than the hard state spectra from 1991 and 1997, but the
fluxes in all energy intervals generally correlate with each other.
The 1999 spectra  show clearly different dependence: the high
energy fluxes are correlated, but there is a clear anti-correlation
between 3--12 keV   and  100--300 keV fluxes due to pivoting.
The timing behaviour has also changed its nature in late 1998,
as  was pointed out by \citet{p03}.

\subsection{QPO frequencies}

\begin{figure}
\centerline{\epsfig{file=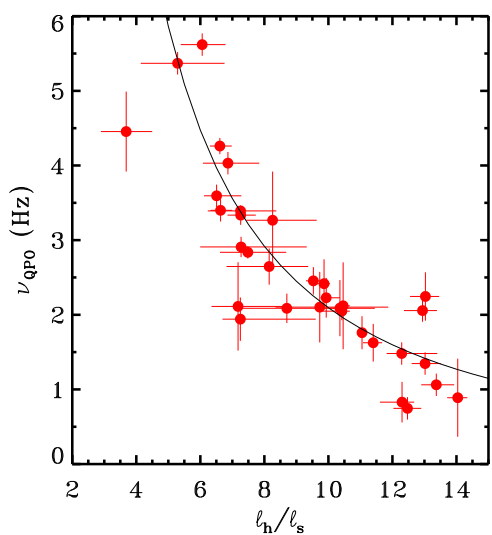,width=8.5cm}}
\caption{Dependence between  QPO frequencies and $\lhls$. The
$\lhls$ are best fit values from model 2. The QPO frequencies are
from \citet{axe05}.  The solid curve shows the best power-law
fit $\nu_\mathrm{QPO}\propto (\lhls)^{-1.48}$. }
\label{fig:timing}
\end{figure}

Among the data sets we have studied, there are
 timing data for 33 observations available from the recent
paper by \citet*{axe05}. This allows us to check the
relation between the characteristic frequencies in the power
density spectra and the spectral parameters. In the work of
\citet{axe05},  several values of  QPO frequencies
$\nu_\mathrm{QPO}$ might be determined during one observation, and
for these data points we assume the averaged middle frequency and
consider the uncertainty  from lower to higher of obtained
values.
 In agreement with
earlier results of GCR99, we find a clear anti-correlation between
the characteristic frequencies of the aperiodic variability and
$\lhls$ (Fig.~\ref{fig:timing}), indicating an apparent connection
between QPO frequencies and the parameters of the Comptonizing
region. This provides an argument (but not a proof) in favour of
the presence of a hot inner corona and a variable inner radius of
the surrounding disc.

The frequency--hardness correlation can be described by a
power-law $\nu_\mathrm{QPO}\propto (\lhls)^{-\alpha}$
with $\alpha =-1.48 \pm 0.04$.
The best fit is shown by the
solid curve on Fig.~\ref{fig:timing}. Similar
correlation was observed by  \citet{p03} and \citet{nov02}.

\subsection{Comparison between phenomenological and
physical spectral models and $R-\Gamma$ correlation}

\label{sec:old_vs_new}

\begin{figure*}
\centerline{\epsfig{file=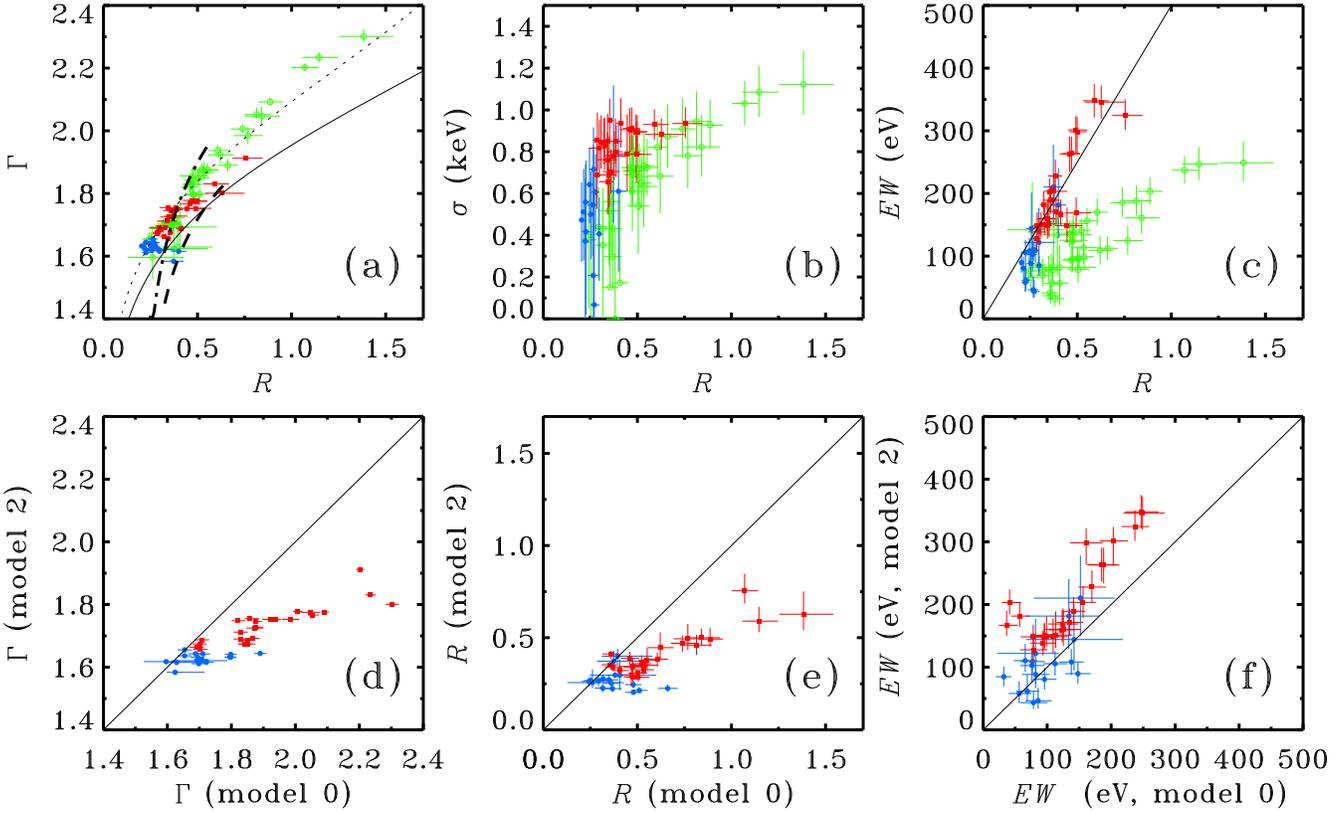,width=\textwidth}} \caption{
(a-c) Correlations obtained using model 0 (green open circles), model 1 (blue
filled circles) and model 2 (red filled squares). (a) The
spectral slope $\Gamma$ vs.  reflection scaling factor $\Refl$
correlation. For model 0, $\Gamma$ is a fitting parameter, for
model 2 -- index of a power-law fitted to the spectral model in
the 2--10 keV range. The solid curve shows the dependence
predicted for the ejection model (\citealt*{bel99}; MBP01) with
the parameters of $i=50\degr$, the albedo of 0.15, $\mu_{\rm
s}=0.5$ and $\tau=2$, the dotted curve -- the same for $\mu_{\rm
s}=0.4$. The dashed curve shows the dependence predicted by the
model with a hot spherical corona and cold overlapping disc
(\citealt{pkr97}; ZLS99) with the black body temperature of 0.2
keV, see Sect.~\ref{sec:old_vs_new} for details; the dot-dashed
curve shows the same model with the dissipation effect taken into
account (dissipation parameter $L_\mathrm{int}= 1$,
$L_\mathrm{int}/4\pi=1$ at $r<1$, see Appendix \ref{sec:append}
for details). (b)  The relativistic smearing Gaussian width
$\sigma$ at 6.4 keV vs.\ $\Refl$. (c)  The equivalent width of the
6.4 keV line $EW$ vs.\ $\Refl$. The straight line is
$EW\mathrm{[eV]}=500R$. (d--f) $\Gamma$, $\Refl$ and $EW$ obtained
from model 2 vs. those from model 0.}
\label{fig:old_vs_new}
\end{figure*}

Fig.~\ref{fig:old_vs_new} compares the results obtained with the
simple phenomenological model of power law + reflection in the
3--20 keV energy range (model 0) with those from our model 2
applied to the 3--1000 keV range. More elaborate and physically
justified models utilizing the full energy range of our data do
not change the picture qualitatively. On the quantitative level we
find that the simple power law + reflection spectral fits to the
3--20 keV data overestimated the amplitude of the reflected
component $\Refl$ and the slope $\Gamma$ of the primary
Comptonization continuum. We confirm, however, that the simple
models did rank correctly the spectra according to the strength of
the reflected component and slope of the Comptonized radiation, as
it was demonstrated in the original publications on this subject
(ZLS99; GCR99). It is also illustrated by the lower three panels
of Fig.~\ref{fig:old_vs_new}.  The difference of the obtained
parameters comes from the fact, that  for wide-energy observations
the main thermal Comptonization component that describes well the
hard energy tail may lie well below the observed flux in the 2--10
keV range (see Fig.~\ref{fig:thprimer} and \ref{fig:nthprimer}) and
has a different  slope in this band. The difference is largest for
1996, 1998 and 1999 data, while in the 1991 and 1997 cases
the soft excess is weak and parameters obtained with the physical models
are similar to those obtained with the phenomenological ones.

On Fig.~\ref{fig:old_vs_new}a, we see a clear correlation between
$\Gamma$ and $\Refl$. For comparison, we also show the
dependencies predicted by the plasma ejection model of
\citet{bel99,bel99b} and MBP01 (cylindrical geometry with
$h/r=2$), using the geometric parameter of that model of
$\mu_\mathrm{s}=0.4$ and 0.5, $i=50\degr$, albedo of the
reflecting medium of 0.15 and $\tau=2$. We used the dependence
between the amplification factor $A$ of Comptonization and
$\Gamma$ from MBP01. We also compared our data with the dependence
expected in the model of ZLS99, assuming the black body
temperature of 0.2 keV, appropriate for Cyg X-1, with one minor
change. In the original paper, all the reflection luminosity was
assumed to reach the observer. Reflection amplitude is an integral
that consists of two parts, from the disc inside the corona and
from the outer part of it. We multiplied the part of luminosity
coming from the former part by $e^{-\tau}$, to approximately take
into account scattering of radiation in the corona ($\tau=1$ was
chosen).  We see that this model cannot quantitatively describe
the presented data. Moreover, taking into account intrinsic
dissipation in the disc (see Appendix \ref{sec:append} for
details) will further increase the slope of the dependence making
the discrepancy larger. Intrinsic dissipation becomes important
for a small inner disc radius (when reflection is relatively
large) and the increase of soft seed photon flux in that case
makes the spectrum softer (see \citealt{bel01}).

\subsection{Physical scenario}

The hard  spectral state of black hole binaries is commonly defined as
the state in which the spectrum is dominated by the hard
Comptonization component, without significant contribution of the
blackbody-type  emission from the optically thick accretion disc.
Naturally, the hard  state is not characterized by a single, uniquely
defined spectrum, but rather includes a continuum of spectral
shapes with the major spectral parameters vary in a rather broad
range.
The diversity of the hard state spectra reflects the position
of the source with respect to the ``bottom hard'' state and the soft
state. Quantitatively, this position can be characterized by the
strength of the reflected component (reflection scaling factor
$\Refl$ or Fe line equivalent width $EW$) or properties of
the main Comptonized component (Comptonization parameter, or Compton
amplification factor $A=\lhls$, or the photon index $\Gamma$ in the
low energy limit) or characteristic frequencies of the aperiodic
variability. Existence of good correlations between all these
quantities suggests that they all are an equally good measure of the
source position within the hard state. The results  found in earlier
work (ZLS99; GCR99; \citealt{gcr00}; \citealt*{gcr04}) and presented in the previous
sections of this paper suggest the following pattern of spectral and
temporal variability. Increase of the strength of the reflected
component is accompanied by the increase of the width of
the Fe line, increase of the characteristic QPO frequencies and
softening of the Comptonized component observed as increase of its
photon index $\Gamma$ in the low energy limit.

We find from our spectral analysis that in the ``bottom hard''
state the broad band spectrum (3-1000 keV) is mostly described by
the single thermal Comptonization spectrum with superimposed
component due to reflection of the primary emission from
relatively cool and neutral, or partly ionized, optically thick
matter (the accretion disc), with an additional relatively weak
soft component.
As the source moves towards the soft state, the strength of the
reflected component increases, and the soft component
becomes more significant.
 Considering the 3--1000 keV energy range
covered by our data, this spectral component reveals itself most
clearly in the $E\la 10$ keV energy domain as the ``soft excess''.
Another independent indication of  spectral complexity
is the gamma-ray power-law tail detected
at MeV energies by the COMPTEL telescope \citep{mccon02}.

From the point of view of the formal fit quality, the $E\la 10$ keV
excess can be described equally well by an additional
thermal Comptonization component due to low temperature, low Comptonization parameter
plasma or by non-thermal Comptonization with the power law
index of accelerated electrons $\Gamma_{\rm inj}\sim 2-3$.
Owing  to the complex shape of the continuum at these energies defined
by the superposition of several spectral components,
these two possibilities can not be easily discriminated based solely
on the low energy data. The task is further complicated by the rather
limited low energy coverage provided by the PCA instrument, $E> 3$
keV. However the above possibilities predict very different behaviour
in the $\sim$ MeV energy domain, where the main thermal Comptonization
component diminishes and the power law tail due to the non-thermal
Comptonization should reveal itself.
As the OSSE sensitivity and energy range are insufficient to probe
existence of the ~MeV tail correlated with the $E<10$ keV excess, we
can not, strictly speaking, give preference to either of these two
models.

There are however several additional considerations to be taken
into account: (1) COMPTEL detected a weak MeV tail in the averaged
hard state data for Cyg X-1 \citep{mccon02}. The slope and
amplitude of this tail is qualitatively consistent with the
extrapolation of the non-thermal Comptonized component, required
to explain the $E\la 10$ keV excess (Fig.\ref{fig:nthprimer}); (2)
in the soft state the non-thermal power law is the dominant (the
only) high energy component; (3) the  $E\la 10$ keV excess is
more pronounced  in the spectra characterized by large reflection
and rather steep slope of the main Comptonized component, i.e. in
those sufficiently close to the soft state. Its strength seems to
increase with increase of the reflection. These arguments suggest
that the non-thermal origin of the $E\la 10$ keV excess is more
plausible.  We note, that the 1991 and 1997 data show much
weaker excess. This may be explained by the lower relative
luminosity of the non-thermal Comptonization component which
therefore reveals itself at lower energies, below the $E=3$ keV
threshold of the PCA instrument, but can be detected by
instruments which have response at lower energies, i.e.   \sax\
(see \citealt{fro01,ds01}).

The overall qualitative picture can be outlined as follows.
The overall geometry of the accretion flow is adequately represented
by the truncated disc model with the inner radius of the standard
optically thick geometrically thin disc varying from $\sim
3$ $R_g$ to $\sim$ several tens $R_g$. Inside this radius the
accretion flow proceeds via quasi-spherical optically thin hot flow.
The plausible mechanism governing the transition from the disc
accretion to the coronal flow is the disc evaporation process as
proposed by \citet{mm94}. The geometrically thin disc gives rise
to the soft black body type component. In addition, due to dynamo,
solar-type magnetic flares  can be produced above the accretion disc \citep{grv79}.
The electrons there can be accelerated and form non-thermal distribution.
Comptonization of the disc emission on these electrons results
in  the power law-like Comptonized emission. The inner optically thin
flow gives rise to the thermal Comptonization component. The relative
contributions of non-thermal and thermal Comptonized components
are defined by the fractions of the gravitational energy released in
the disc (i.e. outside $R_{\rm in}$) and in the inner hot flow
(inside $R_{\rm in}$).
The position of the transition radius is defined by the mass accretion
rate and is modified by the irradiation-related effects. The
transition radius decreases as the mass accretion rate increases.

The QPOs are due to some processes in the transition region near
$R_{\rm in}$ and approximately scale with the Keplerian frequency and
other characteristic time scales of the coronal flow and standard
accretion disc in the transition region.

The configuration with the large inner disc radius, probably
$R_{\rm in}\ga 50-100 R_g$, corresponds to the classical hard
state. The main features of this ``bottom hard'' state are low
strength of the reflected continuum, relatively narrow fluorescent
Fe line of small equivalent width, large Comptonization parameter of
the thermal Comptonized component (hard spectra with the low energy
photon index $\Gamma\sim 1.6$), low frequencies of QPOs. As only small
fraction of the gravitational energy is released in the disc, the
contribution of the non-thermal component is small and the spectrum is
adequately described by thermal Comptonization.

As the mass accretion rate increases, the transition radius decreases,
the disc moves towards the compact object. This results in increase of
the reflection, broader fluorescent Fe line, larger QPO frequencies,
smaller  $\lhls$, i.e. smaller Comptonization parameter in the
inner hot flow. The contribution of the non-thermal component
increases. The optical depth of the thermal plasma of the inner flow
decreases due the shrinking of the inner hot flow as the
disc extends towards the compact object.

The classical soft state (we ignore all the complications and
sub-states here) corresponds to the accretion disc extending all the
way towards the last stable orbit or very close to it. Correspondingly
the inner hot flow disappears and the dominant or the only hard
component is the one due to non-thermal Comptonization of the disc
emission on the non-thermal electrons accelerated in the magnetic
loops/flares above the disc.

The behaviour of the temperature of the thermal Comptonization component is
unclear. It seems relatively constant, which suggests of possible presence of
electron-positron pairs (see MBP01).

The  physical scenario qualitatively outlined above is based on the
truncated disc picture and on the assumption that the spectral
evolution is governed by the change of the transition radius between
the standard accretion disc and the hot inner flow. In this
picture many of the observed
correlations can be explained naturally. However,
the $\Refl-\Gamma$ correlation is significantly better quantitatively
explained by the non-stationary corona  model (MBP01), in which the
governing parameter is the velocity of the blobs of emitting plasma
relative to the accretion disc.
We note that  both models are geometrical in their nature and the
predicted qualitative relations between the physical parameters are
obtained with a number of simplifying assumptions. Therefore
results of quantitative comparison of the model predictions with the
observed pattern of the spectral variability should be interpreted
with caution and any conclusions regarding validity of either model
based on such a comparison would be premature.

\section{Conclusions}

Based on the broad band (3--1000 keV) data from simultaneous
observations by \Ginga\/ and \gro/OSSE in 1991 and
PCA and HEXTE instruments aboard \xte\ and OSSE
in 1996--1999 we studied the spectral
variability of Cyg X-1.

\begin{enumerate}
\item
We confirm earlier results on $\Refl-\Gamma$ correlation.
Considering the 3--20 keV data we find very tight one-parameter
relations between reflection, spectral index and the width of the
Fe line.

\item
More elaborate and physically justified models utilizing the full
energy range of our data do not change the picture qualitatively.
On the quantitative level we find that the simple power law +
reflection spectral fits to the 3--20 keV data overestimated the
amplitude of the reflected component $\Refl$ and the slope
$\Gamma$ of  the primary Comptonization continuum. We confirm,
however, that the simple models did rank correctly the spectra
according to the strength of the reflected component and slope of
the Comptonized radiation, as it was demonstrated in the original
publications on this subject (ZLS99; GCR99).

\item
Based on the analysis of the broad band data we found that
the  spectra in our sample can be adequately described by the
thermal Comptonized component with superposed reflection from the
optically thick disc and a soft excess. This excess is relatively
weak in case of hardest spectra of our sample ($\Gamma\sim1.7$).
As the strength of the reflection increases, the excess becomes
much more significant. Presence of this excess was the primary
reason for the simple spectral approximations of the 3--20 keV
data to overestimate both $\Refl$ and $\Gamma$. The nature of this
excess cannot be unambiguously determined from our data. Based on
the circumstantial evidence we suggest that it is the lower energy
part of the non-thermal Comptonized component with the power law
index of accelerated electrons $\Gamma_{\rm inj}\sim 2-3$. At
higher energies this non-thermal component reveals itself as a
power law detected by COMPTEL at MeV energies in the average hard
state spectrum of Cyg X-1.

\item We note the variability of the absorption
correlated with the phase of binary system.
These results confirm previous findings of the
X-ray dips in the source.

\item The overall pattern of spectral and temporal variability can be
summarized as follows.  Increase of the strength of the reflected
component is accompanied by the increase of the width of
the Fe line, increase of the characteristic QPO frequencies and
softening of the Comptonized component observed as increase of its
photon index $\Gamma$ in the low energy limit or, equivalently,
decrease of the Compton amplification factor
$\lhls$. Simultaneously, the optical depth of the thermal
Comptonization decreases and the fractional contribution of the
non-thermal component to the total energy flux increases.
The exact behaviour of the electron temperature in the hot inner flow is not
constrained by our data.

\item
We suggest a qualitative physical scenario naturally explaining the
observed behaviour. In this scenario the evolution of the spectral
parameters is governed by the value of the transition radius between
the standard optically thick accretion disc and the inner
quasi-spherical hot flow. The thermal Comptonized component originates
in the inner hot flow as a result of Comptonization of the soft
photons emitted by the accretion disc. The origin of the non-thermal
component is related to the optically thick disc, for example it can
be produced due to non-thermal electrons accelerated near the surface
of the optically thick disc in the solar-type magnetic flares.
The relative contributions of non-thermal and thermal components to
the total energy flux depends on the fractions of the gravitational
energy of accreting matter released in the optically thick disc and in
the hot inner flow.

\end{enumerate}

\section*{Acknowledgments}

We are grateful to Bryan Irby (NASA/GSFC) for help with converting
the \Ginga\ data, and to Magnus Axelsson for sharing the
results of the timing analysis.   This work was
supported by the Centre for International Mobility and the
V\"ais\"al\"a foundation (AI), the Academy of Finland grants
201079 and 204600 and the Wihuri Foundation (JP), and the NORDITA
Nordic project in High Energy Astrophysics. AI was also supported
by RFFI 02-02-17174 and presidential program for support of
leading scientific schools NSH-1789.2003.2. AAZ was supported by
KBN grants PBZ-KBN-054/P03/2001, 1P03D01827 and 4T12E04727.

\appendix
\section{Dissipation and attenuation in the disc-hot flow model}
\label{sec:append}
ZLS99 have considered an idealized geometrical model for thermal Comptonization, reprocessing and reflection in an accretion flow consisting of a central hot sphere surrounded by a flat cold disc, see fig.\ 2 in ZLS99. The sphere has a unit radius, and the inner radius of the disc can assume any value, $d$. For $d<1$, there is an overlap between the two components. The hot sphere Comptonizes soft seed photons emitted by the disc. In the original model of ZLS99, the disc reprocesses and reemits only the photons emitted by the sphere incident on the disc.

Here, we generalize that model to include intrinsic dissipation in the cold disc (as expected in an accretion flow). Also, we take into account scattering of the Compton-reflected photons in the hot sphere, which was neglected in ZLS99.  For completeness, we give here the full set of relevant equations, but refer the reader to ZLS99 for details of the derivation.

The hot sphere has a unit luminosity and emits isotropically. The total flux incident on the disc at a radius, $r$, is then given by (ZLS99),
\begin{equation}
F_{\rm inc} (r) = {3h(r) \over 16\upi^2},
\end{equation}
where
\begin{eqnarray}
\lefteqn{ h(r)=(4/3)\times} \nonumber\\ \lefteqn{
\cases{ \left[\left(2-r^{-2}\right)E\left(r^2\right)+
\left(r^{-2}-1\right)K\left(r^2\right)\right], & $r<1$,\cr
\left[\left(2r-r^{-1}\right)E\left(r^{-2}\right)+
2\left(r^{-1}-r\right)K\left(r^{-2}
\right)\right], & $r\geq 1$,\cr } }
\label{hr}
\end{eqnarray}
where $E$ and $K$ are complete elliptic integrals. The luminosity of the disc due to reemission of photons incident on it is
\begin{equation}
L_{\rm inc}(d)=4\upi \int_d^\infty {\rm d}r\, r F_{\rm inc}(r).
\end{equation}
This can be divided into the contributions to the integral from the parts of the disc at $d\leq 1$ and $d>1$, $L_{\rm inc}=L_{\rm inc}^{<}+L_{\rm inc}^{>}$, where the first term is nonzero only for $d<1$.

Then, the relative strength of Compton reflection can be identified (ZLS99) with the ratio of the luminosity of the disc due to irradiation to the fraction of the sphere luminosity that is  not incident on the disc. However, we correct here for attenuation of the reflection from the parts of the disc with $d<1$ because of scattering by the hot electrons in the sphere. For the radial optical depth of the sphere  $\tau$, we can then write
\begin{equation}
R\simeq {{\rm e}^{-\tau} L_{\rm inc}^{<} + L_{\rm inc}^{>} \over 1- L_{\rm inc}}.
\end{equation}

Another effect not included in the treatment of ZLS99 is the intrinsic dissipation in the disc. Far away from the center (so any effect of the inner boundary condition is negligible), the dissipated flux per unit area is $\propto r^{-3}$, and we assume it for $r>1$. On the other hand, the dissipation in the part of the disc inside the hot sphere, $r<1$, is reduced due to the transfer of the power to the hot plasma. We assume here that dissipation either to be null or constant matching that of the outside disk,
\begin{equation}
F_{\rm int}(r)={L_{\rm int} \over 4\upi} \cases{
0\quad {\rm or}\quad 1, & $r\leq 1$,\cr r^{-3}, & $r>1$,\cr}
\end{equation}
where we parametrized the relative intrinsic dissipation by the dimensionless factor, $L_{\rm int}$, defined as the intrinsic luminosity of the disc extending from $r=1$ to infinity (regardless of the actual value of $d$).

The power in seed photons scattered in the sphere from both the reprocessing and the intrinsic dissipation (assuming $\tau=1$) is then
\begin{eqnarray}
\lefteqn{
L_{\rm s}(d) \approx {3(1-a) \over 4\upi^2} \int_d^\infty {\rm d}r\, r h^2(r)
+{L_{\rm int}\over \upi} \int_{\max(d,1)}^\infty {\rm d}r\, {h(r)\over r^2} }\nonumber\\
\lefteqn{\qquad
+{L_{\rm int}\over \upi} \int_d^1 {\rm d}r\, r h(r), }
\label{Ls}
\end{eqnarray}
where $a$ is the albedo and the term given by the second line of equation (\ref{Ls}) appears only in the case of the nonzero dissipation within the sphere and when $d<1$.

The amplification factor of the process of thermal Comptonization is then
$A(d)\equiv 1/L_{\rm s}(d)$. It can be related to the spectral index of the power-law part of the Comptonization spectrum using e.g. the formula of MBP01
\begin{equation}
\Gamma=C (A-1)^{-\delta},
\end{equation}
where $C=2.19$ and $\delta=0.14$ for BHB with $\ktseed=0.2$ keV.

\label{lastpage}
\end{document}